\documentclass[journal=jpccck,manuscript=article,layout=twocolumn]{achemso}

\setkeys{acs}{
abbreviations=true,
articletitle=true,
biblabel=plain,
chaptertitle=true,
doi=true,
email=true,
etalmode=truncate,
keywords=true,
maxauthors=100,
super=true
}

\setcitestyle{super,open={},close={}}

\usepackage[utf8]{inputenc} 
\usepackage{textcomp} 
\usepackage[T1]{fontenc} 
\usepackage[portuguese,english]{babel} 
\usepackage[version=4]{mhchem} 
\usepackage{siunitx} 
\usepackage{graphicx} 
\graphicspath{{}} 
\usepackage{geometry} 
\usepackage[format=plain,justification=justified,singlelinecheck=false,font={normalsize,bf},labelfont=bf,labelsep=space]{caption} 
\usepackage{float} 
\usepackage{natbib} 
\usepackage{setspace} 
\usepackage{xkeyval} 
\usepackage{array} 
\usepackage{booktabs} 
\usepackage{listings} 
\usepackage{lmodern} 
\usepackage{mathpazo} 
\usepackage[breaklinks,colorlinks=true,allcolors=blue]{hyperref} 
\hypersetup{breaklinks,colorlinks=true,linkcolor=blue,filecolor=blue,urlcolor=blue,citecolor=blue}
\hypersetup{colorlinks=true,citecolor=blue,urlcolor=blue}
\newcommand{\doi}[1]{\href{http://dx.doi.org/#1}{\nolinkurl{#1}}}
\usepackage{subfigure} 
\usepackage[compact]{titlesec} 
\usepackage{natmove} 
\usepackage{multirow} 
\usepackage{lipsum} 
\usepackage{calc} 
\usepackage{longtable} 
\usepackage{tabularx} 
\usepackage{placeins} 
\usepackage{tablefootnote}
\usepackage{adjustbox}
\usepackage{titletoc}
\usepackage{latexsym}
\usepackage{chemformula}
\usepackage{cancel}
\usepackage{amsmath}
\usepackage{amssymb}
\usepackage{amsfonts}
\usepackage{amstext}
\usepackage{xr}
\usepackage{cleveref}
\graphicspath{{Figures/}}

\usepackage[none]{hyphenat}
\usepackage{microtype}
\makeatletter
\let\l@addto@macro\relax
\makeatother
\usepackage[fontsize=12pt]{scrextend}

\SectionsOn
\SectionNumbersOn
\AbstractOn

\title[TITLE]{Unveiling the Impact of Organic Cation Passivation on Structural and Optoelectronic Properties of Two-Dimensional Perovskites Thin Films}

\author{Israel C. Ribeiro}
\affiliation[USP]{S{\~a}o Carlos Institute of Chemistry, University of S{\~a}o Paulo, P.O. Box $780$, $13560$-$970$, S{\~a}o Carlos, SP, Brazil}

\author{Pedro Ivo R. Moraes}
\affiliation[USP]{S{\~a}o Carlos Institute of Chemistry, University of S{\~a}o Paulo, P.O. Box $780$, $13560$-$970$, S{\~a}o Carlos, SP, Brazil}

\author{Albert F. B. Bittencourt}
\affiliation[USP]{S{\~a}o Carlos Institute of Chemistry, University of S{\~a}o Paulo, P.O. Box $780$, $13560$-$970$, S{\~a}o Carlos, SP, Brazil}

\author{Juarez L. F. Da{~}Silva}
\email{juarez_dasilva@iqsc.usp.br}
\affiliation[USP]{S{\~a}o Carlos Institute of Chemistry, University of S{\~a}o Paulo, P.O. Box $780$, $13560$-$970$, S{\~a}o Carlos, SP, Brazil}


\usepackage[utf8]{inputenc}
\DeclareUnicodeCharacter{2212}{-}

\begin{document}


\begin{tocentry}
    \centering
    \includegraphics[width=0.90\linewidth]{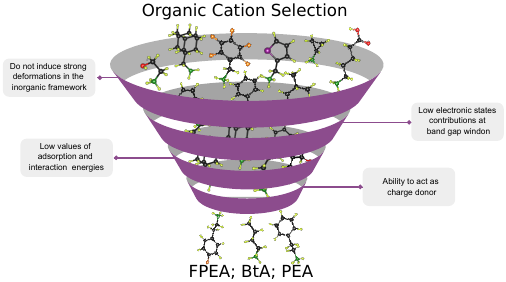}
\end{tocentry}

\begin{abstract}
Several organic cations have been used to passivate perovskite films; however, selecting the optimal cation remains challenging. In this work, we carried out density functional theory calculations to understand the effects induced by \num{17} different organic cations on the passivation ($P$-cations) of thin two-dimensional \ce{$P$_{2}($MA$_{$n$–1})Pb_{$n$}I_{$3n+1$}} perovskites films, where $n = 1$ and $2$. We found that the interactions between different types of $P$-cations and the inorganic slab affect the length and angles of the bonds within the inorganic framework (\ce{PbI6}-octahedra). In general, the binding mechanism includes the interactions of organic cations with the inorganic framework, which leads to the accumulation of electron density within the halides, indicative of Br{\o}nsted--Lowry acid-base interactions. Oxygenated groups facilitate additional \ce{H}-bond formation through \ce{\bond{-}OH} and \ce{\bond{-}COOH} groups, promoting the localization of the electron density between layers and improving the energetic stability of the system. Based on the results and analysis, we found that three $P$-cations might have higher potential for real-life applications, namely 4-fluorophenylethylammonium (FPEA), phenylethylammonium (PEA) and butylammonium (BtA).
\end{abstract}

\section{Introduction}

Hybrid organic-inorganic metal halide perovskites have emerged as a prominent class of materials for perovskite solar cell (PSC) applications.\cite{Kojima_6050_2009,Gao_2206387_2023} For example, the power conversion efficiency (PCE) has risen from \SI{3.8}{\percent}\cite{Kojima_6050_2009} in \num{2009} to about \SI{26}{\percent}\cite{Zheng_1153_2024} in \num{2024}, that is, increasing steadily along \num{15} years. Their remarkable potential is attributed to their excellent optoelectronic characteristics, including a high absorption coefficient,\cite{DeWolf_1035_2014,Park_65_2015} low exciton binding energy,\cite{Zhang_4505_2013,Ponseca_5189_2014} and outstanding charge transport mobility.\cite{Stoumpos_9019_2013}

The photoactive layer in PSCs is formed by three-dimensional (3D) perovskites with the general chemical formula \ce{$ABX$3}, in which $A$ represents an organic or inorganic monovalent cation such as methylammonium (\ce{MA}), formamidinium (\ce{FA}), \ce{Cs+}, and so on, while $B$ is a divalent cation (e.g., \ce{Pb^{2+}}, \ce{Sn^{2+}}), and $X$ is a halide anion (\ce{I-}, \ce{Cl-}, \ce{Br-}).\cite{Lekina_189_2019,Paritmongkol_5592_2019} As a result, the possible chemical combinations of $A$, $B$, and $X$ can lead to the formation of large variety of compounds, allowing for precise adjustment of their structural and optoelectronic properties through different chemical configurations.\cite{Danelon_8469_2024}

The widespread application of PSCs faces challenges due to structural stability issues. For example, moisture,\cite{Zhang_2023} ultraviolet irradiation,\cite{Sun_8682_2017} oxygen,\cite{Chi_0897_2021} and heat,\cite{AdilAfroz_2432_2020} are the main environmental factors that affect the stability of real-life applications. To address this issue, numerous strategies have been suggested and thoroughly investigated using experimental or theoretical techniques,\cite{Stoumpos_9019_2013,Wang_1803753_2018,dosSantos_5259_2023,Ribeiro_13667_2023,Danelon_8469_2024} including the use of particular organic molecules as passivation agents in perovskite thin films.\cite{Stoumpos_9019_2013,Hu_171_2019} The passivation process aims to protect perovskite thin films from environmental conditions and stabilize unstable perovskite phases.\cite{Bi_1400_2017,Li_105237_2020}

For example, bulky organic cations with large carbon chains have received increasing attention as surface passivation agents ($P$-cations) with the aim of improving the long-term stability and performance of PSCs.\cite{Zhang_105_2018,ThrithamarasseryGangadharan_2860_2019,Lee_100759_2021} Furthermore, during the passivation process of thin films of \ce{MAPbI3}, large organic molecules can undergo a cation exchange with \ce{PbI2} on the surface of the 3D film, forming a thin atomic two-dimensional (2D) perovskite capping layer. 

The number of inorganic layers ($n$) in perovskite thin films, is highly sensitive to the type of organic molecules used as passivators and the precursors involved in the perovskite thin film synthesis.\cite{Zhang_105_2018} For example, monovalent organic cation (\ce{$P$}-cation) passivation in \ce{MAPbI3} thin films can induce the formation of structures described by the following general chemical formula, \ce{$P$_{2}($MA$_{$n$–1})Pb_{$n$}I_{$3n+1$}}, which is one of the building blocks in the 2D Ruddlesden--Popper perovskite structures.\cite{Ruddlesden_538_1957}

The vast availability of $P$-cations enables the rapid advancement of organic-inorganic optoelectronic devices, where the material properties are highly tunable by the nature of the organic molecule. The dipole of $P$-cations can play a crucial role in modulating the work function and is also related to the mobility of the holes and the performance of the device.\cite{GlvezRueda_824_2020} However, bulky organic cations increase the dielectric confinement and exciton binding energy of the material, thus charge separation and transport are dependent on organic ammonium cations and their supramolecular chemistry.\cite{Zhou_1901566_2019,GlvezRueda_824_2020}  

Li \textit{et al.},\cite{Li_e202217910_2023} successfully addressed the problem of poor carrier transport properties in perovskites films by incorporating $\gamma$-aminobutyric acid (GABA, \ce{C4H10O2N}) as a possible passivation agent. The addition of GABA reduces the exciton binding energy to \SI{73}{\milli\electronvolt} in \ce{(GABA)2MA3Pb4I13}, in comparison to the compound \ce{(BtA)2MA3Pb4I13} (approximately \SI{119}{\milli\electronvolt}), where BtA represents the butylammonium molecule (\ce{C4H12N}), and results in devices with a PCE of approximately \SI{18.73}{\percent}. Incorporation of unsaturated organic passivator improves the carrier pathway and solar cell efficiency, for example, phenylethylammonium (PEA, \ce{C8H12N}) (\SI{15.46}{\percent}) compared to butylammonium (\SI{11.71}{\percent}).\cite{Zhou_1901566_2019} Furthermore, 2D films of perovskites also found applications in X-ray detectors,\cite{Tsai_eaay0815_2020} where the dark current is minimized by depositing a thin layer of \ce{(PEA)2PbI4} onto a pristine \ce{MoS2} photodetector.\cite{Wang_1901402_2020} 

Aliphatic $P$-cations such as BtA have shown remarkable properties, such as low steric hindrance, which facilitate the formation of perovskite films. However, these $P$-cations exhibit insulating properties that accentuate the effects of quantum and dielectric confinement.\cite{Stoumpos_2852_2016,Wang_123589_2020} In contrast, aromatic $P$-cations can offer additional advantages to enhance charge transport in PSCs. These advantages encompass higher dielectric constants, which reduce the dielectric mismatch between organic and inorganic layers.\cite{Qin_1904050_2020} Aromatic $P$-cations, through \ce{\pi\bond{...}\pi} interactions and hydrogen bonding (\ce{H}-bonds), can also contribute to structural stability and improved molecular packing rearrangement. This contrasts to aliphatic cations with softer organic tails that provide greater flexibility and weaker interactions during film formation.\cite{Cohen_2588_2019,Lao_2105307_2022} 

In this work, we present an extensive theoretical investigation, based on density functional theory calculations, to explore the effects induced by the passivation of 2D perovskite thin films by different $P$-cations. The selected perovskites can be represented by the \ce{$P$_{2}($MA$_{$n$–1})Pb_{$n$}I_{$3n+1$}} chemical formula, where $n = 1$ represents slabs composed by a single \ce{PbI6}-octahedra monolayer and $n = 2$ indicates slabs composed by two \ce{PbI6}-octahedra layers. For $n = 2$, we have selected $MA$ (methylammonium), while considering a set of \num{17} $P$-cations (aryl, aromatic and with oxygenated groups (\ce{\bond{-}OH} and \ce{\bond{-}COOH})) to act as passivation agents.

Our findings reveal that $P$-cations significantly affect the structural and optoelectronic properties of the materials. The binding mechanism involves charge transfer between the organic and inorganic thin film, which is expected and indicating a Br{\o}nsted--Lowry acid-base behavior. Furthermore, oxygenated groups can form additional \ce{H}-bonds, which enhance energy stability and impact the electronic band gap. The optimal choice of $P$-cation depends on the intended application; longer alkylammonium chains contribute to improved stability, while functionalization of alcohol/carboxylic acid enhances energetic stability and charge transfer. Furthermore, aromatic $P$-cations offer stability through conjugation, which regulates the crystal arrangement. Fluorinated $P$-cations have an impact on energy stability, with the \ce{F} positions influencing structural stability and the properties of the charge carrier. Therefore, we selected three $P$-cations with high potential for PSC applications: 4-fluorophenylethylammonium (FPEA), phenylethylammonium (PEA) and butylammonium (BtA).

\section{Theoretical Approach and Computational Details}

\subsection{Total Energy Calculations}

Our calculations are based on the DFT\cite{Hohenberg_B864_1964,Kohn_A1133_1965} framework within the semilocal Perdew--Burke--Ernzerhof\cite{Perdew_3865_1996} (PBE) formulation for the exchange-correlation (XC) energy functional, as implemented in the Vienna \textit{ab{~}initio} simulation package (VASP),\cite{Kresse_558_1993,Kresse_11169_1996} version $5.4.4$, where the KS orbitals are expanded in plane waves. We used the all-electron projector augmented wave (PAW) method\cite{Blochl_17953_1994,Kresse_1758_1999} to describe the interaction between the core and the valence electrons. Semilocal XC functionals have been known to have limitations in providing an accurate description of long-range nonlocal van der Waals (vdW) interactions, e.g. adsorption of molecular species on surfaces.\cite{Qin_1904050_2020,Moraes_16357_2023} Thus, we used the semi-empirical D3 vdW correction\cite{Grimme_154104_2010} to improve the description of selected monovalent cations on the \ce{MAPbI3} thin films.\cite{Zhang_105_2018,Ozorio_3439_2020,Li_e202217910_2023}

The selected monovalent $P$-cations have different geometries and molecular sizes, which can affect the magnitude of the inplane ($xy$) equilibrium lattice parameters ($a_0, b_0$). Therefore, to account for these effects, all equilibrium structures were obtained by optimizing the inplane stress tensor and atomic forces in all directions ($xyz$),\cite{Francis_4395_1990} using a plane wave cutoff energy of $1.50{\times}\text{ENMAX}_{max}$, where $\text{ENMAX}_{max}$ is the maximum recommended cutoff energy within PAW projectors for each particular compound. For example, \SI{651.647}{\electronvolt} (compounds with \ce{O}), \SI{731.547}{\electronvolt} (compounds with \ce{F}), and \SI{631.353}{\electronvolt} for the remaining compounds. Throughout the stress-tensor calculations, the magnitude of the stress tensor along the $z$-direction was not optimized (repeated slab geometry), however, all atomic forces were optimized without constraints.

For the remaining properties, such as binding energies, density of states (DOS), band structures, work function, and net atomic charges computed using the density-derived electrostatic and chemical (DDEC) method,\cite{Manz_47771_2016,Limas_45727_2016} we employed a plane wave cutoff energy of \SI{548.660}{\electronvolt} for all systems. For the integration of the Brillouin zone (BZ), we employed \textbf{k}-mesh of $2{\times}2{\times}1$ for the optimization of the structure and $4{\times}4{\times}1$ for the remaining properties. All equilibrium structures were obtained once the atomic forces in each atom were smaller than \SI{2.50e-2}{\electronvolt\per\angstrom} and with a total energy convergence of \SI{e-5}{\electronvolt}. 
                             
\subsection{Selection of Organic Monovalent Cations as Passivation Agents}

There are a huge number of possible organic monovalent cations that can act as passivation agents for perovskite thin films.\cite{Marchenko_7383_2020,Moral_2001431_2020} Few descriptors have been suggested to optimize the selection process of monovalent $P$-cations as passivation agents, which include: $(i)$ the atomic radius of the $P$-cations should be larger than the typical $A$-cations commonly used in \ce{$ABX$3} compounds, i.e., large than \SI{2.6}{\angstrom} (Goldschmidt tolerance factor);\cite{Lee_100759_2021} $(ii)$ should induce small structural deformations in the inorganic framework;\cite{Zhang_1154_2020} $(iii)$ should have the ability to act as charge donor to the thin film, e.g., a net positive charge in the amine group, in which preferring cations with primary ammonium over secondary, etc., in descending order.\cite{Hou_060906_2020}

\begin{figure}[t!]
    \centering
    \includegraphics[width=0.80\linewidth]{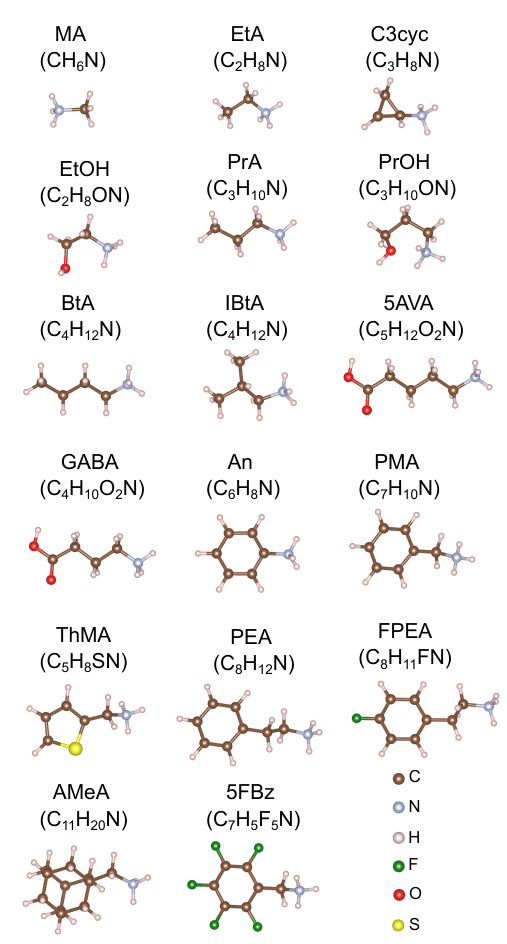}
    \caption{Molecular representation of the selected monovalent $P$-cations in the optimized neutral configurations. The chemical formulas are enclosed within parentheses for clarity. The abbreivations are defined within the text.}
    \label{fig:organic_bindings}
\end{figure}

Based on the mentioned rules, we selected a set of \num{17} $P$-cations, which are shown in Figure{~}\ref{fig:organic_bindings}, namely, methylammonium (MA, \ce{CH6N}), ethylammonium (EtA, \ce{C2H8N}), phenylethylammonium (PEA, \ce{C8H12N}), propan-1-aminium (PrA, \ce{C3H10N}), butylammonium (BtA, \ce{C4H12N}), isobutylammonium (IBtA, \ce{C4H12N}), anilinium (An, \ce{C6H8N}), phenylmethylammonium (PMA, \ce{C7H10N}), 1-adamantylmethylazanium (AMeA, \ce{C11H20N}), cyclopropylazanium (C3cyc, \ce{C3H8N}), 4-fluorophenylethylammonium (FPEA, \ce{C8H11FN}), perfluorobenzylammonium (5FBz, \ce{C7H5F5N}), ethanolamine (EtOH, \ce{C2H8ON}), 3-aminopropan-1-ol (PrOH, \ce{C3H10ON}), 5-valeric ammonium acid (5AVA, \ce{C5H12O2N}), 2-thiophenemethylammonium (ThMA, \ce{C5H8SN}), and $\gamma$-amino butyric acid (GABA, \ce{C4H10O2N}). 

\subsection{Two-Dimensional Perovskite Thin Films}

Primarily, the 2D thin films of perovskites structures were conceptually obtained by cutting along the crystallographic $(100)$ of the 3D \ce{MAPbI3} orthorhombic ($2{\times}2{\times}2$) supercell  and adding a vacuum thickness of \SI{15}{\angstrom}, i.e., the well known repeated slab geometry approximation.\cite{DaSilva_703_2006} The slabs are characterized by the general chemical formula \ce{$P$_{2}($MA$_{$n$–1})Pb_{$n$}I_{$3n+1$}}, where $n = 1$ and $n = 2$ corresponds to a inorganic monolayer and an inorganic bilayer. As mentioned a few times, the $P$-cations act as passivators for the inorganic layers (\ce{PbI6}-octahedra). 

\begin{figure*}[t!]
    \centering
    \includegraphics[width=0.80\linewidth]{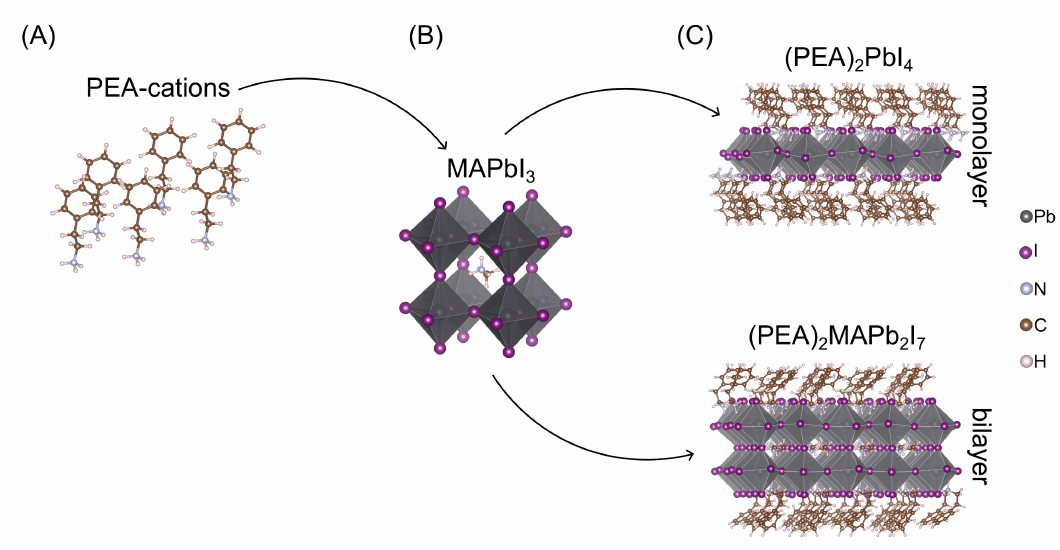}
    \caption{Passivation model via organic $P$-cations, using PEA-cations as example (A), on \ce{MAPbI3} (B) with formation of monolayer and bilayer 2D perovskites thin films (C).}
    \label{fig:passivation_procedure_2dpvk}
\end{figure*}

Therefore, according to the electron counting rule, each side of the slab (top and bottom) requires \num{4} $P$-cations for a complete surface passivation, resulting in the \ce{$P$8Pb4I16} ($n = 1$, 1L) and \ce{$P$8(MA)4Pb8I28} ($n = 2$, 2L) models. It is important to note that these chemical formulas represent the number of chemical species in our models, rather than the unit formulas of the compounds. Additional details are illustrated in Figure{~}\ref{fig:passivation_procedure_2dpvk}.

\subsection{Passivation on Perovskites Layers} 

Regarding the conformation of organic \ce{$P$}-cations, using the \ce{MA} molecule as a reference for the construction of other organic bindings, one of the hydrogen attached to carbon was enumerated by organic groups such as alkyl (EtA, PrA, BtA), alicyclic (C3cyc), tricyclic (AMeA) and aromatic (PMA, PEA, ThMA, 5FBz). The presence of a rotatable methylene ammonium \ce{R\bond{-}NH3} head group in bulk organic cations has been observed to result in the formation of both the ferroelectric phase (Cmc21) and the paraelectric phase (Cmca), which in turn gives rise to ordered and disordered 2D orientations in hybrid organic-inorganic perovskite (HOIP) systems.\cite{Hou_060906_2020} 

\begin{figure}[t!]
    \centering
    \includegraphics[width=0.80\linewidth]{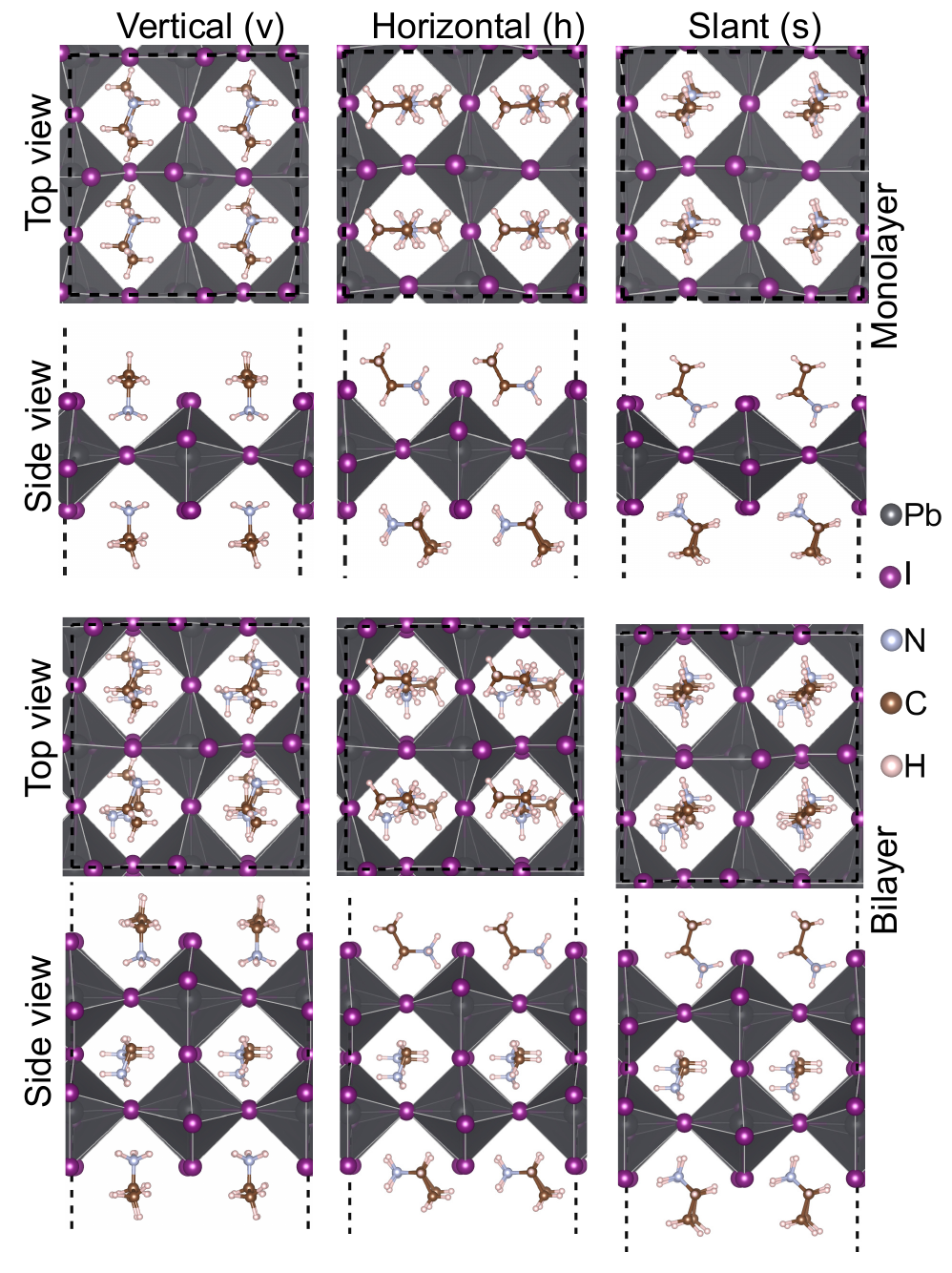}
    \caption{Top and side views of the vertical ($v$), horizontal ($h$), and slant ($s$) orientations of the \ce{R\bond{-}NH3} group in relation to the \ce{PbI2} plane, for the case of \ce{EtA_{2}PbI4} (monolayer) and \ce{EtA_{2}MAPb2I7} (bilayer) perovskites. The dashed line represents the unit cell.}
    \label{fig:top_side_configurations}
\end{figure}

Here, we considered the \ce{R\bond{-}NH3} oriented in the vertical ($v$), horizontal ($h$), and slanted ($s$) relative to the plane \ce{PbI2} to examine the dipole effects of the interface of head group orientations, as illustrated in Figure{~}\ref{fig:top_side_configurations}. Compared to DFT-PBE+D3 calculations performed within our group, geometric optimization using stress tensor and atomic forces for these systems was a challenge. In the following, we summarize the steps performed:
\begin{enumerate}
    \item We started the optimization considering the inplane lattice parameters ($a_{0}$, $b_{0}$) frozen, while $c_0$ was kept fixed along all calculations. Through the selective dynamics feature implemented in VASP, we kept the inorganic slab frozen, allowing only the $P$-cations to move. This step was carried out using only the $\Gamma$-point for BZ integration and a plane wave cutoff energy of $0.875{\times}\text{ENMAX}_{max}$. We performed an average of \num{300} ionic steps for each $P$-cation to ajust to the substrate. 

    \item In the second step, we optimize the atomic forces on all atoms within the unit cell using the same computational parameters as in the first step.

    \item In the $3rd$ step, we optimized the inplane ($xy$) stress tensor and the atomic forces on all atoms. For that, we performed only the \num{25} steps, that is, the first changes in the parameters $a_0$ and $b_0$. In this step, we increase the cutoff energy to $1.250{\times}\text{ENMAX}_{max}$ and use the $\Gamma$-point for the BZ integration.
    
    \item In the $4th$ step, we optimized only the atomic forces (\num{200} ionic steps) on all atoms using a cutoff energy of $1.125{\times}\text{ENMAX}_{max}$ and only the $\Gamma$-point for the integration of the BZ. 

    \item In the $5th$ step, we optimize the inplane stress tensor and atomic forces on all atoms considering a plane-wave cutoff energy of $1.500{\times}\text{ENMAX}_{max}$ and a \textbf{k}-mesh of $2{\times}2{\times}1$ for the BZ integration. At the end, we obtain the equilibrium parameters $a_0$ and $b_0$ for all configurations.

    \item Finally, using the equilibrium lattice parameters $(a_0, b_0)$, we optimize the atomic forces on all atoms without constraints using a plane wave cutoff energy of $1.125{\times}\text{ENMAX}_{\text{max}}$ and a \textbf{k}-mesh of $2{\times}2{\times}1$ for the BZ integration.
\end{enumerate}

\section{Results and Discussion}

\begin{figure*}[t!]
    \centering
    \includegraphics[width=0.80\linewidth]{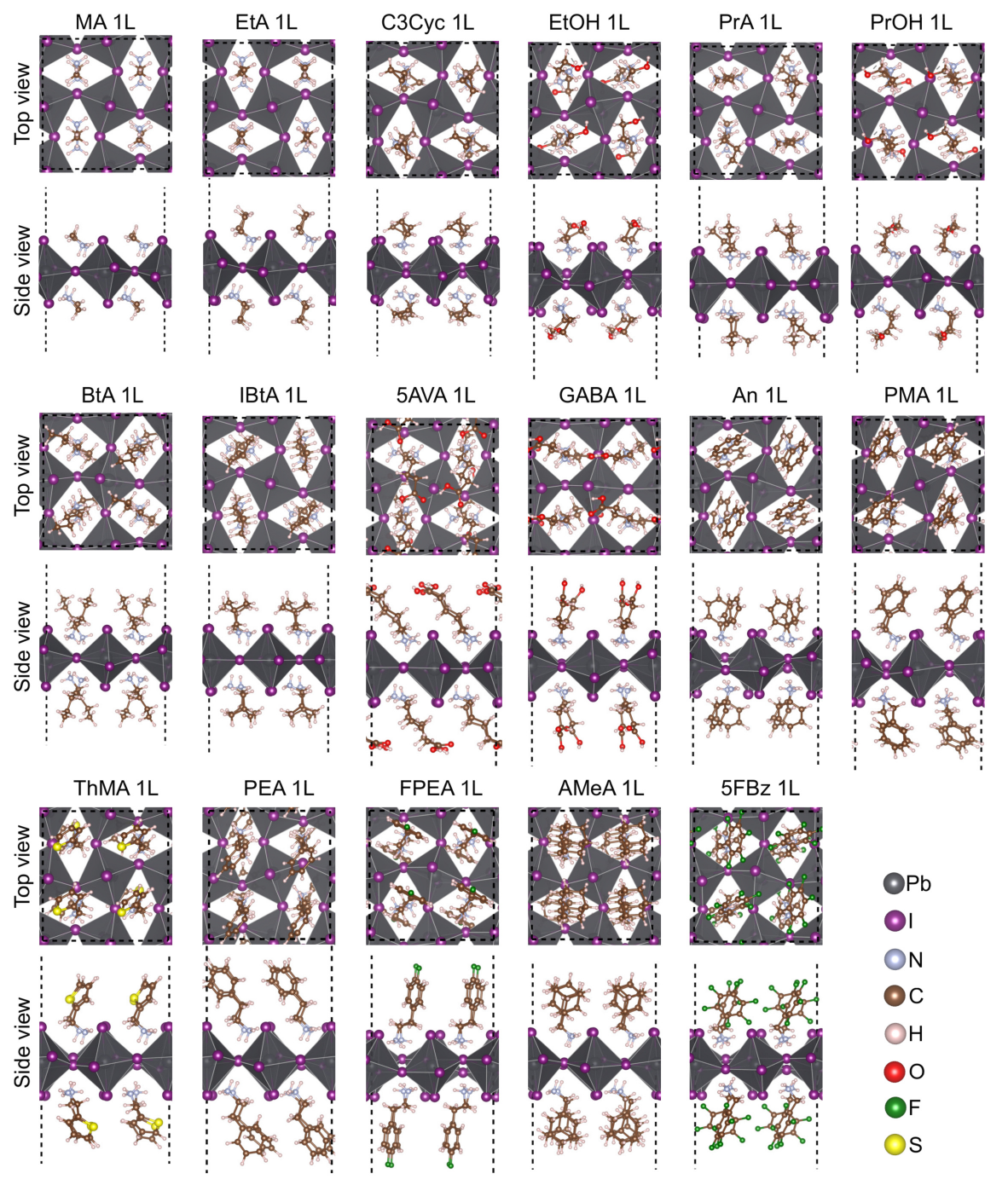}
    \caption{Top and side views of the monolayers 2D perovskites thin films, which have the \ce{$P$2PbI4} compositions ($P$ = $P$-cations). The dashed line represents the unit cell.}
    \label{fig:monolayers_structures}
\end{figure*}

\begin{figure*}[t!]
    \centering
    \includegraphics[width=0.80\linewidth]{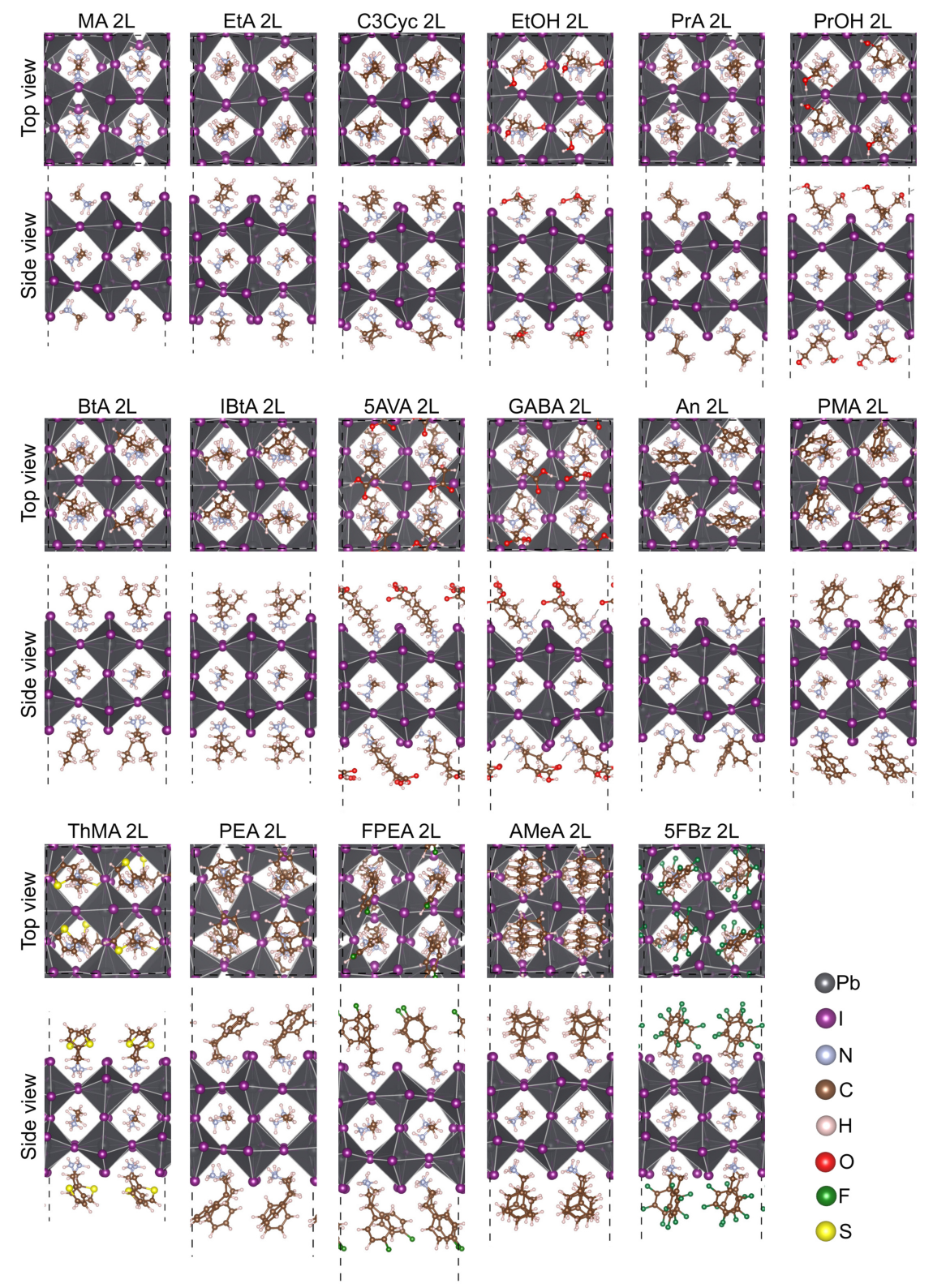}
    \caption{Top and side views of the bilayers 2D perovskites thin films, which have the \ce{$P$2MAPb2I7} compositions ($P$ = $P$-cations). The dashed line represents the unit cell.}
    \label{fig:bilayers_structures}
\end{figure*}

Several structural optimizations were performed, producing the lowest energy configurations for each respective system, as depicted in Figure{~}\ref{fig:monolayers_structures} and Figure{~}\ref{fig:bilayers_structures}. To improve our atomistic understanding of the passivation effects induced by organic monocations (\ce{R\bond{-}NH3+}), we performed structural, energetic, electronic, and optical characterizations using several descriptors (physicochemical properties). In the subsequent discussion, we will highlight the most significant findings, while additional data and analyses are provided within the Electronic Supporting Information (SI) file. 

\subsection{Preferential $P$-Cation Orientation on Perovskite Slabs}

The \ce{R\bond{-}NH3} moiety plays a crucial role in 2D perovskites from the optoelectronic point of view.\cite{Zhang_1154_2020} To investigate its role, we performed several structural optimizations for different conformations, depending on vertical ($v$), horizontal ($h$), and slant ($s$) orientations of the \ce{R\bond{-}NH3} group relative to the plane \ce{PbI2} formed after passivation of each $P$-cation, as illustrated in Figure{~}\ref{fig:top_side_configurations}. According to our results, after passivation, there was a consistent reduction in \ce{C\bond{-}N} ($d_{av}^{\ce{C\bond{-}NH3}}$) distances, evident in monolayer and bilayer films. The values converged to an average of \SI{1.49}{\angstrom}, emphasizing the influence of interactions between the $P$-cation and the inorganic slab. 

Furthermore, the transition from the gas phase of the $P$ -cations molecules to the passivated phase induced a significant variation in the molecular conformation, especially in the presence of larger organic chains and functional groups, leading to novel intermolecular interactions. It is crucial to note that the structures generated with C3cyc and An as $P$-cations exhibited steric hindrance for the $h$ orientations; therefore, such configurations were not considered.

\begin{figure}[t!]
    \centering
    \includegraphics[width=0.80\linewidth]{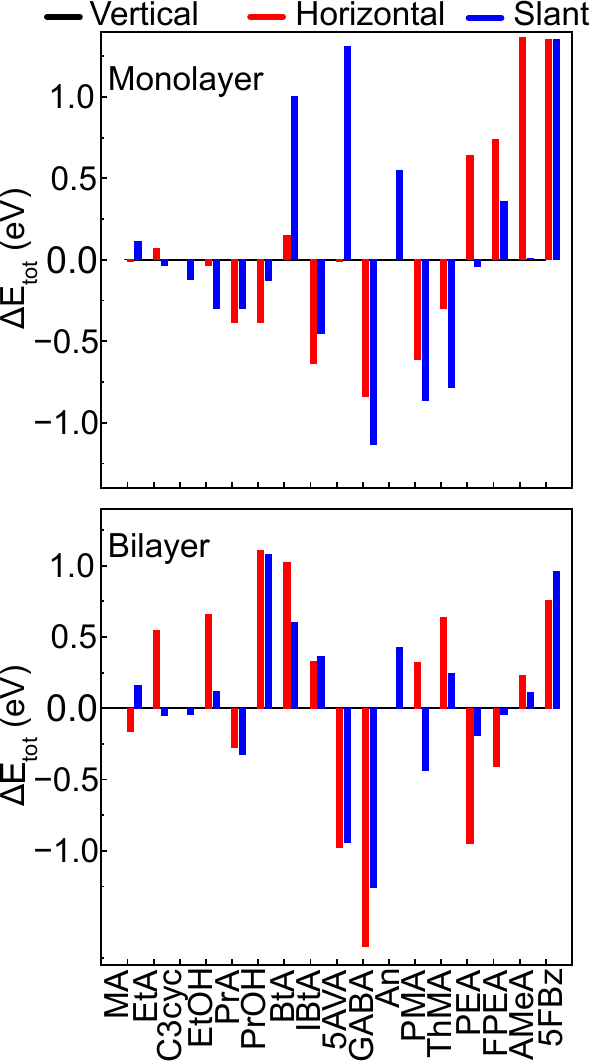}
    \caption{Relative total energy (${\Delta}E_{tot}$) for all configurations ($v$, $h$, and $s$) for both monolayer and bilayer perovskite structures. Here, the $v$ orientations are at ${\Delta}E_{tot}$ = \num{0}.}
    \label{fig:relative_energy}
\end{figure}

For the relative energy differences (${\Delta}E_{tot}$), we choose the $v$ orientations of each system as the reference energy system, which is critical to obtain reliable comparison among the different molecular configurations. According to the ${\Delta}E_{tot}$ results shown in Figure{~}\ref{fig:relative_energy}, \SI{41}{\percent} of the monolayers conformations have lower energy for the $s$ configuration, whereas for the inorganic bilayer slabs, \SI{47}{\percent} of the conformations have lower energy for the $v$ configuration. Some structural parameters, such as hydrogen bonds (\ce{H}-bonds), could clarify the energy stability of compounds. 

\subsection{Formation of \ce{H}-bonds on the $P$-Cation/Slab Interface} 

\begin{figure}[t!]
    \centering 
    \includegraphics[width=0.80\linewidth]{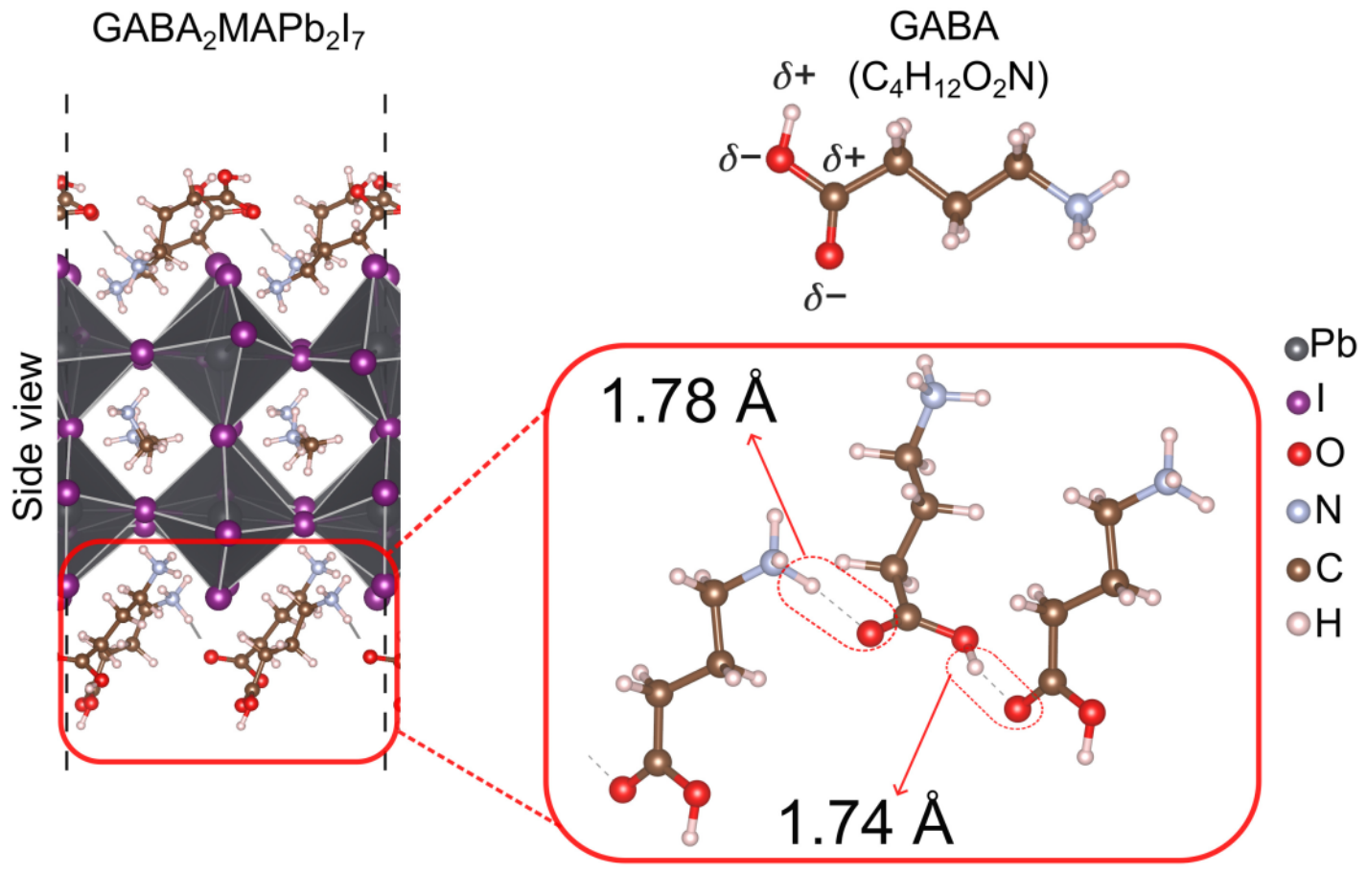}
    \caption{Molecular representation of the formation of intermolecular interactions of the \ce{H}-bonds type, as well as the length of the \ce{H}-bonds via \ce{OH\bond{...}O} and \ce{NH\bond{...}O} in bilayer GABA systems.}
    \label{fig:intermolecular_interactions}
\end{figure}

Previous studies have claimed a direct correlation between energetic stability and the formation of a \ce{H}-bond in perovskite-based materials,\cite{Cheng_1_2018,Ozorio_2286_2021,Leung_1_2022} and therefore it is important to characterize the formation of \ce{H}-bonds for the studied systems. In all 2D systems examined, we found shorter and directional hydrogen-iodide bond distances ($d^{\ce{H\bond{...}I}}$) for the \ce{NH\bond{...}I} bond. We established a cutoff radius of \SI{3.0}{\angstrom} for $d^{\ce{H\bond{...}I}}$, which is logically within the hydrogen bonding range ($\approx \SI{3.0}{\angstrom}$).\cite{Espinosa_170_1998} 

When analyzing the configurations $v$, $h$, and $s$, in general, it is observed that the behavior of $d_{av}^{\ce{H\bond{...}I}}$ decreases for structures with a lower total energy. Consequently, it can be inferred that the \ce{H}-bond length is directly correlated with the energetic stability of the system. Therefore, for further discussion, we will consider the characterizations of the lowest energy structures between the $h$, $v$, and $s$ configurations for each $P$-cations. Thus, we have \num{17} monolayer and \num{17} bilayer structures, represented respectively by Figure{~}\ref{fig:monolayers_structures} and Figure{~}\ref{fig:bilayers_structures}.

The $d_{av}^{\ce{H\bond{...}I}}$ values differ between 1L and 2L for structures with the same \ce{R\bond{-}NH3} group orientation, suggesting that the quantum confinement effect can affect the molecular conformation of the $P$-cation. The AMeA $P$-cation is characterized by a substantial number of hydrogen atoms. Therefore, this system demonstrates a notable relative disparity in $d_{av}^{\ce{H\bond{...}I}}$ between the 1L and 2L configurations, which is approximately \SI{12.05}{\percent}. We observed lower values for $P$-cations with fewer hydrogens, e.g. EtA at \SI{0.38}{\percent}.

For structures that contain $P$-cations with oxygenated groups, \ce{\bond{-}COOH} and \ce{\bond{-}OH} in their tails, we found \ce{H}-bonds through hydrogen-oxygen interactions such as \ce{OH\bond{...}O}, \ce{CH\bond{...}O}, and \ce{NH\bond{...}O} with the neighboring $P$-cation, Figure{~}\ref{fig:intermolecular_interactions}. These extra polar interactions play a crucial role in decreasing exciton binding energy and improving the PCE of PSCs.\cite{Cheng_1_2018}

The \ce{OH\bond{...}O} interactions present in structures passivated with molecules such as GABA, 5AVA, EtOH, and PrOH play a crucial role in optoelectronic properties, due to the ability to promote charge transfer between adjacent ligands in the structure. In addition, the type of \ce{H}-bonds, $d^{\ce{OH\bond{...}O}}$, effectively localizes charges in the interlayer region close to the \ce{PbI6}-octahedra, playing a crucial role in the binding mechanism process of the 2D perovskites.\cite{Cheng_1_2018,Li_e202217910_2023} 

\subsection{Inplane Slab Lattice Deformations} 

\begin{figure*}[t!]
    \centering
    \includegraphics[width=0.80\linewidth]{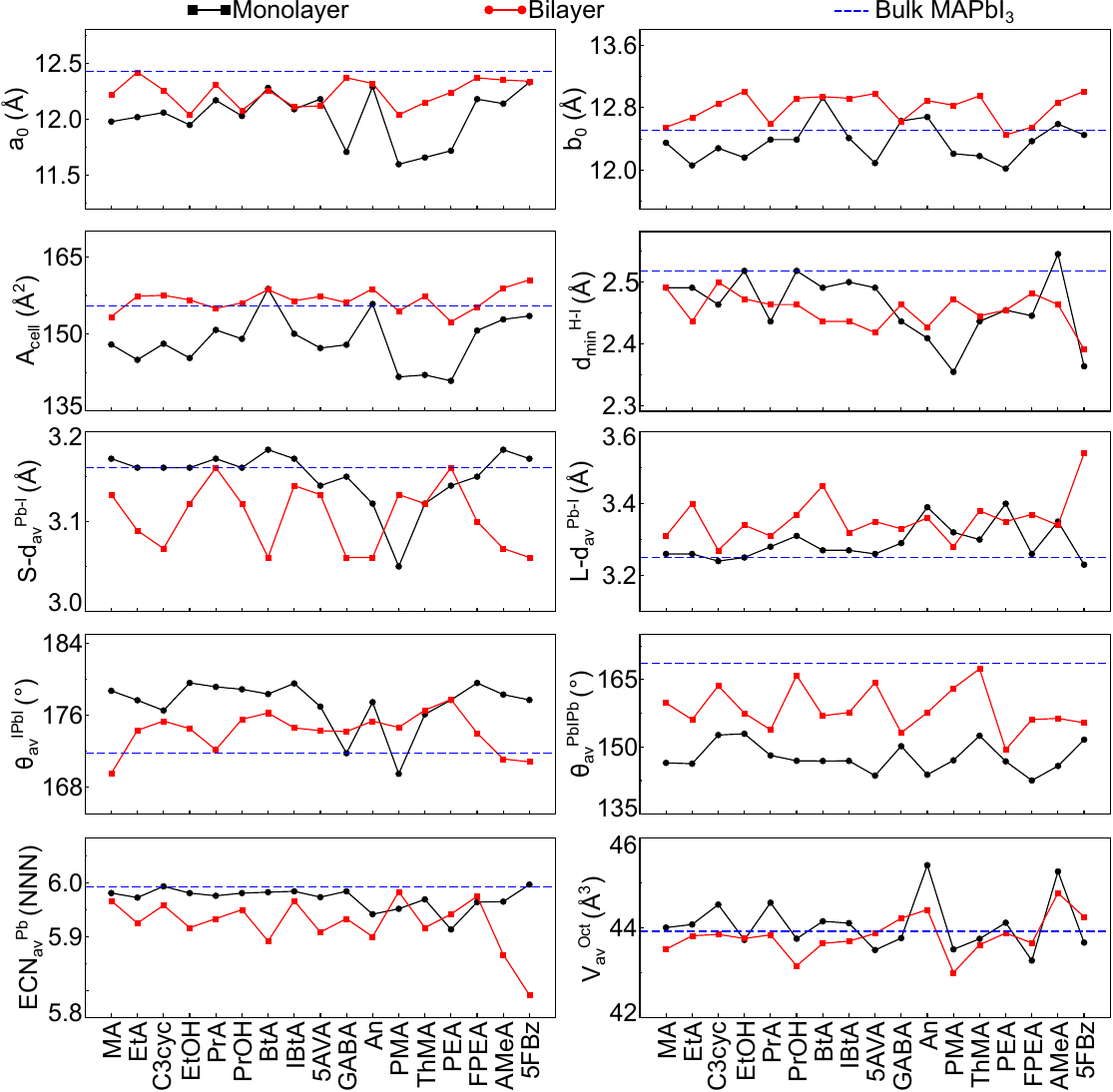}
    \caption{Equilibrium structural parameters for 2D \ce{$P$2PbI4} and \ce{$P$2(MA)Pb2I7} systems: lattice parameters ($a_{0}$, $b_{0}$), surface unit cell area ($A_{cell}$), the average hydrogen bond distance with the iodine ($d_{av}^{\ce{H\bond{...}I}}$), average bond distance of the short (S-$d_{av}^{\ce{Pb\bond{-}I}}$) and long (L-$d_{av}^{\ce{Pb\bond{-}I}}$) bonds, average angles for the \ce{I\bond{-}Pb\bond{-}I} ($\theta_{av}^{\ce{IPbI}}$) and \ce{Pb\bond{-}I\bond{-}Pb} ($\theta_{av}^{\ce{PbIPb}}$) combinations, the effective coordination number for the \ce{Pb} (ECN$_{av}^{\ce{Pb}}$) and the average volume of the \ce{PbI6}-octahedra ($V_{av}^{Oct}$).}
    \label{fig:structural_properties}
\end{figure*}

To characterize the role of $P$-cations in the slab lattice, we selected two structural descriptors, namely: $(i)$ equilibrium lattice parameters ($a_0$, $b_0$) and $(ii)$ surface area ($A_{cell}$). All results are shown in Figure{~}\ref{fig:structural_properties}. Monolayer systems have lower equilibrium lattice constants ($a_{0}$, $b_{0}$) than the corresponding bilayer systems as a result of the quantum confinement effect. Furthermore, \ce{PbI6}-octahedra distortions, which result from out-of-phase distortions and short and long \ce{Pb\bond{-}I} bonds, have a considerable impact on $a_0$ and $b_0$; these results are consistent with previous studies.\cite{Zhang_1154_2020,Ribeiro_13667_2023}  

However, the lattice constants can also be influenced by the $P$-cation size and intermolecular forces, e.g. \ce{H}-bonds shown in Figure{~}\ref{fig:structural_properties}. In addition to the inherent characteristics of the crystal lattice, the size of the $P$-cation can affect the arrangement of the atoms within the lattice. When the $P$-cation is larger in size, it can introduce strain and distort the regularity of the lattice. In contrast, a smaller $P$-cation can lead to a more compact and tightly packed lattice structure. 

In terms of the lattice properties of monolayer systems, we found a pattern for straight-chain $P$-cations. First, the values of $a_{0}$ decrease as the linear chain increases, i.e., MA < EtA < PrA < BtA. These differences in the equilibirum lattice parameters of systems induced by the nature of the carbon chain can have an effect on the thermal properties of 2D perovskites.\cite{Cuthriell_11710_2023} In addition, when comparing $P$-cations with the same number of carbons, we also found a decrease in $a_{0}$ for branched-type cations, e.g. \ce{BtA \bond{->} IBtA}, as well as for both 1L and 2L when considering the following films: \ce{EtA \bond{->} EtOH}; \ce{PrA \bond{->} PrOH}; \ce{BtA \bond{->} GABA}. This relation may be associated with the formation of \ce{H}-bonds in systems passivated with \ce{\bond{-}OH} and \ce{\bond{-}COOH}.

For passivation with aromatic $P$-cations, the \ce{\pi \bond{...} \pi} interactions can affect the lattice, and our data indicate that there is a decrease in lattice parameters as the branch bonded to the aryl group increases, thus following the sequence \ce{An \bond{->} PMA \bond{->} PEA}. As a result, they exhibit a regulated arrangement and orientation of 2D perovskites, which can be seen from the side view Figure{~}\ref{fig:monolayers_structures}, and thus according to data reported for these types of $P$-cations.\cite{Gao_1_2023}

Finally, we define the area of the unit cell of the surface ($A_{cell}$), as the product $a_0$ $\times$ $b_0$. In this context, $A_{cell}$ is significantly influenced by the effects induced on the lattice parameters. Consequently, bilayer systems exhibit higher $A_{cell}$ values compared to their corresponding monolayers. This can be attributed to the reduced quantum confinement effect and also the role of passivation of the $P$-cation and the presence of MA interlayers, which collectively contribute to the observed increase in surface area.

\subsection{\ce{PbI6}-Octahedra Distortions}

For a deeper structural characterization, the following local geometric descriptors were selected: $(i)$ average bond lengths for the \ce{PbI6}-octahedra, $(ii)$ bond angles, $(iii)$ average effective coordination numbers for \ce{Pb} (ECN$_{av}^{\ce{Pb}}$), and $(iv)$ average volume of \ce{PbI6}-octahedra ($V_{av}^{oct}$). All results are shown in Figure{~}\ref{fig:structural_properties}, and the data are discussed in more detail below.

Due to the electrostatic interactions between the $P$-cation charges and the negative charge on the surface of the thin films, the \ce{Pb\bond{-}I} bond distances along the \ce{PbI6}-octahedra are closely related. However, according to a hexacoordinated metal complex, we have axial and equatorial bonds that may differ when interacting with ligands.\cite{Temerova_19220_2022} As a result, each octahedron contains short (S-$d_{av}^{\ce{Pb\bond{-}I}}$) and long (L-$d_{av}^{\ce{Pb\bond{-}I}}$) chemical bonds, Figure{~}\ref{fig:structural_properties}, which contribute to lower distortions along the lattice. 

Intriguingly, we found stronger effects in 2L systems in terms of S-$d_{av}^{\ce{Pb\bond{-}I}}$ and L-$d_{av}^{\ce{Pb\bond{-}I}}$, resulting in overall systems with the lowest, S-$d_{av}^{\ce{Pb\bond{-}I}}$, and highest, L-$d_{av}^{\ce{Pb\bond{-}I}}$, bond length compared to equivalent 1L systems. This pattern has been observed in 2D \ce{BA_{2}(MA_{$n$–1})Pb_{$n$}I_{$3n+1$}} crystals, where reduced S-$d_{av}^{\ce{Pb\bond{-}I}}$ and increased L-$d_{av}^{\ce{Pb\bond{-}I}}$ values of $n = 2$ to $4$ have been discovered.\cite{Stoumpos_2852_2016} In particular, the bilayer 5FBz system stands out with the lowest S-$d_{av}^{\ce{Pb\bond{-}I}}$ and the highest L-$d_{av}^{\ce{Pb\bond{-}I}}$ among the investigated systems, which can be attributed to repulsion between iodine and fluorine halides present in the $P$-cations.

This effect on \ce{Pb\bond{-}I} bonds in 2L systems indicates a relationship to the effects of interlayer cations (MA), which present different \ce{R\bond{-}NH3} group orientations with respect to $P$-cation in 2D material. As stated previously, this effect decreases when $n = 1$, but is minimal for the bulk \ce{MAPbI3} since only MA is present as a cation within the thin film. Therefore, we can assume that for $1 < n < \infty$, there is a mixture of distortion effects in 2D perovskites that arises from the competition of distortions from the $c$-axis and $ab$-plane caused by $P$-cations and MA interlayer molecules, respectively. It is critical to note that the corner-connected octahedra in 1L structures must always rotate in the opposite direction (out of phase) to maintain structural integrity, as seen in the top view in Figure{~}\ref{fig:monolayers_structures}. From the side view, all monolayers show approximately $\SI{0}{\degree}$ tilting.

To gain a comprehensive understanding of octahedral distortions, we also measured the angles formed by the \ce{Pb\bond{-}I\bond{-}Pb} bonds ($\theta_{av}^{\ce{PbIPb}}$) as well as for \ce{I\bond{-}Pb\bond{-}I} ($\theta_{av}^{\ce{IPbI}}$). Monolayer systems, on average, exhibit substantial differences in the values of $\theta_{av}^{\ce{PbIPb}}$ and $\theta_{av}^{\ce{IPbI}}$ compared to the bulk angles, indicating the effect of the number of layers on the distortions. The fact that $\theta_{av}^{\ce{PbIPb}}$ often has lower values than $\theta_{av}^{\ce{IPbI}}$ suggests that out-of-phase distortions account for most of the observed results.

In general, average ECN$_{av}^{\ce{Pb}}$ values for monolayer systems closely resemble those found for bulk \ce{MAPbI3} (ECN$_{av}^{\ce{Pb}} = \SI{5.99}{NNN}$, where \si{NNN} denotes the number of nearest neighbors). Therefore, \ce{PbI6}-octahedra do not present relevant Jahn--Teller type distortions,\cite{JahnH._220_1937} explained by the full electronic shell of lead $d$-states. However, small changes in lead coordination can be attributed to passivation effects, producing distortions and small displacements of the metal center due to \ce{N\bond{-}H\bond{...}I\bond{-}Pb} interactions.

Remarkably, the 5FBz bilayer system displayed the lowest ECN$_{av}^{\ce{Pb}}$ value, about \SI{4.17}{\percent} smaller compared to the bulk, which might be related to distortions of the center cation displacement (CD) caused by highly electronegative ions (\ce{F}) present in the $P$-cation. The average volumes of the \ce{PbI6}-octahedra ($V_{av}^{Oct}$) found in the thin films are relatively close to the value found in the bulk \ce{MAPbI3}. In fact, small variations in $V_{av}^{Oct}$ are observed due to the effects of organic molecules on the \ce{Pb\bond{-}I} bonds. In this way, the structural properties of 2D perovskites have a strong dependence on the physicochemical properties of the $P$-cation used as a passivator.

\subsection{Adsorption and Interaction Energies of $P$-Cation on Perovskite Slabs}

\begin{figure}[t!]
   \centering
   \includegraphics[width=0.80\linewidth]{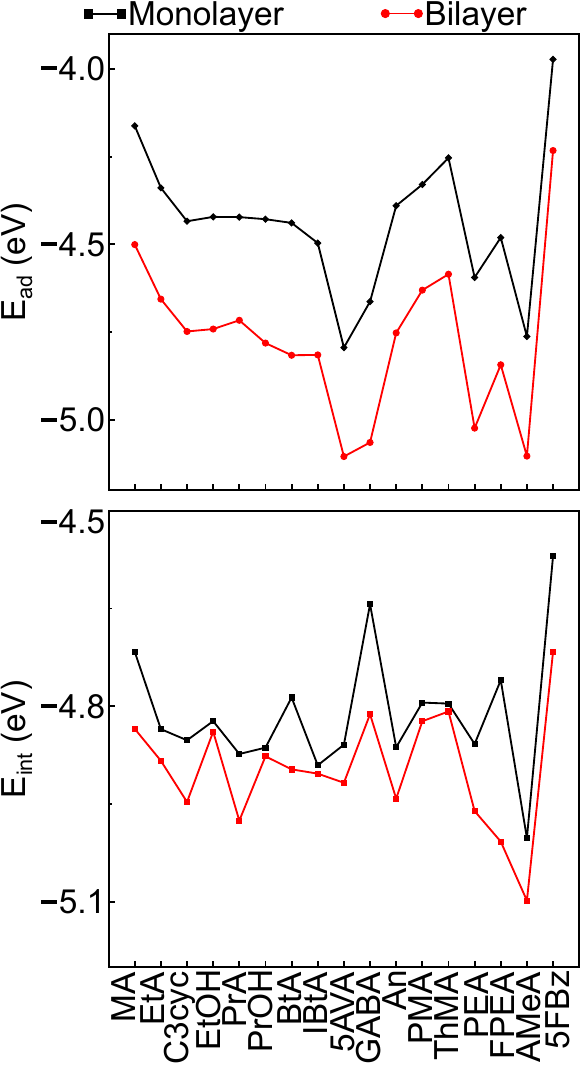}
   \caption{Adsorption ($E_{ad}$) and interaction ($E_{int}$) energies for $P$-cations molecules on perovskite monolayer and bilayer structures.}
   \label{fig:Ead_Eint_energy}
\end{figure}

To assess the energy stability of the $P$-cation under passivation in thin perovskites films, we calculated the energies of adsorption ($E_{ad}$) and interaction ($E_{int}$) for all materials. $E_{ad}$ can be computed using the following equation,
\begin{align}\label{eq:Ead_energy}
    E_{ad} = (E_{tot}^{2D~perovskite} - E_{tot}^{slab~relaxed} & \nonumber \\ - 8\times E_{tot}^{A-cation~free})/8{~},
\end{align}
where $E_{tot}^{2D~perovskite}$ is the total energy of the 2D perovskite, while $E_{tot}^{slab{~}relaxed}$ is the total energy of the optimized perovskite slabs through the stress-tensor without the \num{8} $P$-cations adsorbates, the $E_{tot}^{A-cation~free}$ is the total energy of the free $P$-cation and \num{8} is the number of $P$-cations removed from the slab. For $E_{int}$, the following equation was used,
\begin{align}\label{eq:Eint_first}
    E_{int} = (E_{tot}^{2D~perovskite} - E_{tot}^{slab{~}frozen} & \nonumber \\ - E_{tot}^{A-cation{~}frozen})/8{~},
\end{align}
where $E_{tot}^{slab{~}frozen}$ and $E_{tot}^{A-cation{~}frozen}$ are, respectively, the total energies of the frozen perovskite slab and the \num{8} $P$-cation frozen in their optimized geometric positions. The results $E_{ad}$ and $E_{int}$ are shown in Figure{~}\ref{fig:Ead_Eint_energy}.

The adsorption and interaction of $P$-cations in perovskite slabs are favorable and exhibit significant energetic stability for all systems, including both monolayers and bilayers. This is evident from the magnitude of the values $E_{ad}$ and $E_{int}$, compared to those reported for the polar and nonpolar adsorption of several molecules on the \ce{MoS2} surface.\cite{Regis_107710_2024} This can be explained by the attractive Coulomb interaction between the opposite charges of the $P$-cations and the inorganic slab. Nevertheless, the findings suggest that bilayer systems are more energetically stable than those of the respective monolayers, possibly because of the quantum confinement effect. 

For most systems, the 2D organic component of the $P$-cations provides the highest energetic contribution to the 2D perovskites, resulting in a smaller difference between the $E_{int}$ values of the monolayer and bilayer systems. When comparing passivated systems with oxygenated $P$-cations with those without oxygen, we observe a modest reduction in $E_{int}$. Furthermore, for the system passivated with 5FBz, lower values of $E_{int}$ and $E_{ad}$ are obtained, possibly due to the repulsive effect between the \ce{F-} and \ce{I-} species.

The physicochemical descriptors that characterize organic molecules can be used to monitor the adsorption of $P$-cations, suggesting that the molecular structure of $P$-cations contributes to better adsorption in thin films. Molecules capable of forming more \ce{H}-bonds are energetically favorable, exhibiting higher negative $E_{ad}$ values. According to the findings, $P$-cations with oxygenated groups and $\pi$-bonds had lower values of $E_{ad}$, possibly related to their increased stability in the gas phase compared to other organic structures.

With respect to $E_{int}$, the molecular chemistry of the $P$-cations does not appear to be a determining factor. Similar behavior is observed between the monolayers and their corresponding bilayers, leading to the conclusion that the interaction forces between the $P$-cation and the slab are generally similar between 1L and 2L. The AMeA structure has the lowest $E_{ad}$ and $E_{int}$ values among all systems, which may be related to a molecular structure with a higher number of \ce{H} atoms, energetically favored by \ce{H}-bonds.

\subsection{Charge Transfer via Effective Charges Analysis}

\begin{figure*}[t!]
    \centering
    \includegraphics[width=0.90\linewidth]{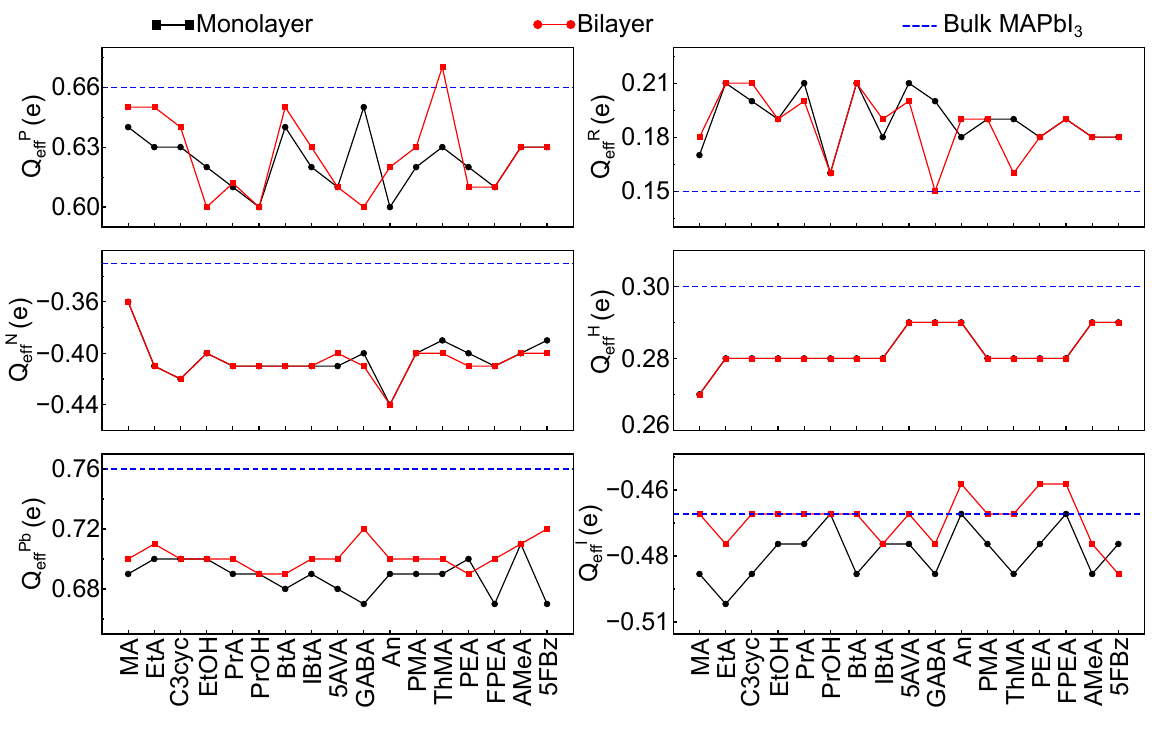}
    \caption{Effective DDEC charges for the $P$-cation, radical $R$, nitrogen and hydrogen both from \ce{NH3} group, lead and iodine, which are indicated by $Q^{P}_{eff}$, $Q^{R}_{eff}$, $Q^{\ce{N}}_{eff}$, $Q^{\ce{H}}_{eff}$, $Q^{\ce{Pb}}_{eff}$ and $Q^{\ce{I}}_{eff}$, respectively.}
    \label{fig:charge_analysis}
\end{figure*}

To enhance our atomistic understanding, we calculated the magnitude of the charge transfer using the density derived electrostatic and chemical (DDEC) method.\cite{Manz_47771_2016,Limas_45727_2016} To evaluate the magnitudes of the charges in the systems, we conducted a local charge analysis investigating the magnitudes of the charges for specific moieties within the full system. Therefore, we determined the average charges of the $P$-cation ($Q^{P}_{eff}$), the radical $R$ bonded to the \ce{R\bond{-}NH3} group ($Q^{R}_{eff}$), nitrogen ($Q^{\ce{N}}_{eff}$), the hydrogen bonded to the \ce{N} ($Q^{\ce{H}}_{eff}$), lead ($Q^{\ce{Pb}}_{eff}$), and iodine ($Q^{\ce{I}}_{eff}$). The results are visually presented in Figure{~}\ref{fig:charge_analysis}, with additional details available in the SI.

The charge transfer process emerges from the interaction effect between the $P$-cation (acid = proton donor) and the halides (base = proton acceptor) through Br{\o}nsted--Lowry acid-base behavior, leading to charge transfer occurrence. In this way, $Q^{P}_{eff}$ can offer us useful information on the charge mechanism. More precisely, our data support a charge transfer mechanism through  \ce{N\bond{-}H\bond{...}I\bond{-}Pb} interactions, where the magnitude depends on the type of $P$-cation. 

The presence of electronegative groups in the carbon chains of $P$-cations improves the polarity of bulk organic compounds, as illustrated in Figure{~}\ref{fig:intermolecular_interactions} for the GABA molecule. This directly influences the $Q^{P}_{eff}$ values and, consequently, the charger transfer process. As mentioned previously, the presence of oxygenated groups induces \ce{H}-bond interactions while simultaneously reducing the van der Waals gap, resulting in improved optoelectronic properties. This includes the effective localization of charges within the $P$-cations regions, facilitating the formation of "charged bridges".\cite{Li_e202217910_2023} This phenomenon, in turn, promotes the overlap of charge densities in adjacent inorganic layers, thereby enhancing the efficiency of charge transfer optimization between these layers.

Due to the resonance electronic characteristics provided by $\pi$-bonds, which can alter charge donation to the inorganic slab, the aromatic $P$-cation passivator exhibits charge transfer dynamics.\cite{Gao_1_2023} The $Q^{P}_{eff}$ deviations for those cations can be explained by the presence of withdrawing groups or electron density donors. When groups with stronger electron density donor properties are introduced, the Br{\o}nsted--Lowry acid character of the $P$-cation increases. This leads to a slight increase in $Q^{P}_{eff}$; for example, we found an increase of about \SI{5}{\percent} in $Q^{P}_{eff}$ when comparing the effects for the \ce{An2PbI4} and \ce{PEA2PbI4} films.

The calculated charge values for radicals $R$ ($Q^{R}_{eff}$), which affect the polarity of the cation and subsequently the charge transfer process, can also be used to explain the values of $Q^{P}_{eff}$. As mentioned above, radicals with highly electronegative groups may form localized interlayer charges. Therefore, due to the inductive effect produced by molecular dipoles, lower levels of $Q^{R}_{eff}$ might result in higher $Q^{P}_{eff}$ values.

Due to the quantum confinement effects present in 2D systems, there is an impact on the results for the charge properties of the inorganic slabs. First, the \ce{Pb} ($Q^{\ce{Pb}}_{eff}$) reveal lower values for the 2D system compared to the bulk system. Furthermore, as observed for \ce{5FBz2PbI4}, the $P$-cation 5FBz strongly influences the \ce{Pb\bond{-}I} bond distances, and as result, we can relate these distortions to the lowest value of $Q^{\ce{Pb}}_{eff}$ found for this film.

However, our results support the notion that the number of inorganic slabs increases the negative charge on the halides ($Q^{\ce{I}}_{eff}$) with values close to $Q^{\ce{I}}_{eff}$ found for the halides in the bulk \ce{MAPbI3} structure. This shows that systems with $n > 1$ have the greatest potential for photovoltaic applications. This observation is consistent with the findings of the existing literature.\cite{Mitzi_1473_1995} 

\subsection{Electron Density Difference}

\begin{figure*}[t!]
    \centering
    \includegraphics[width=0.80\linewidth]{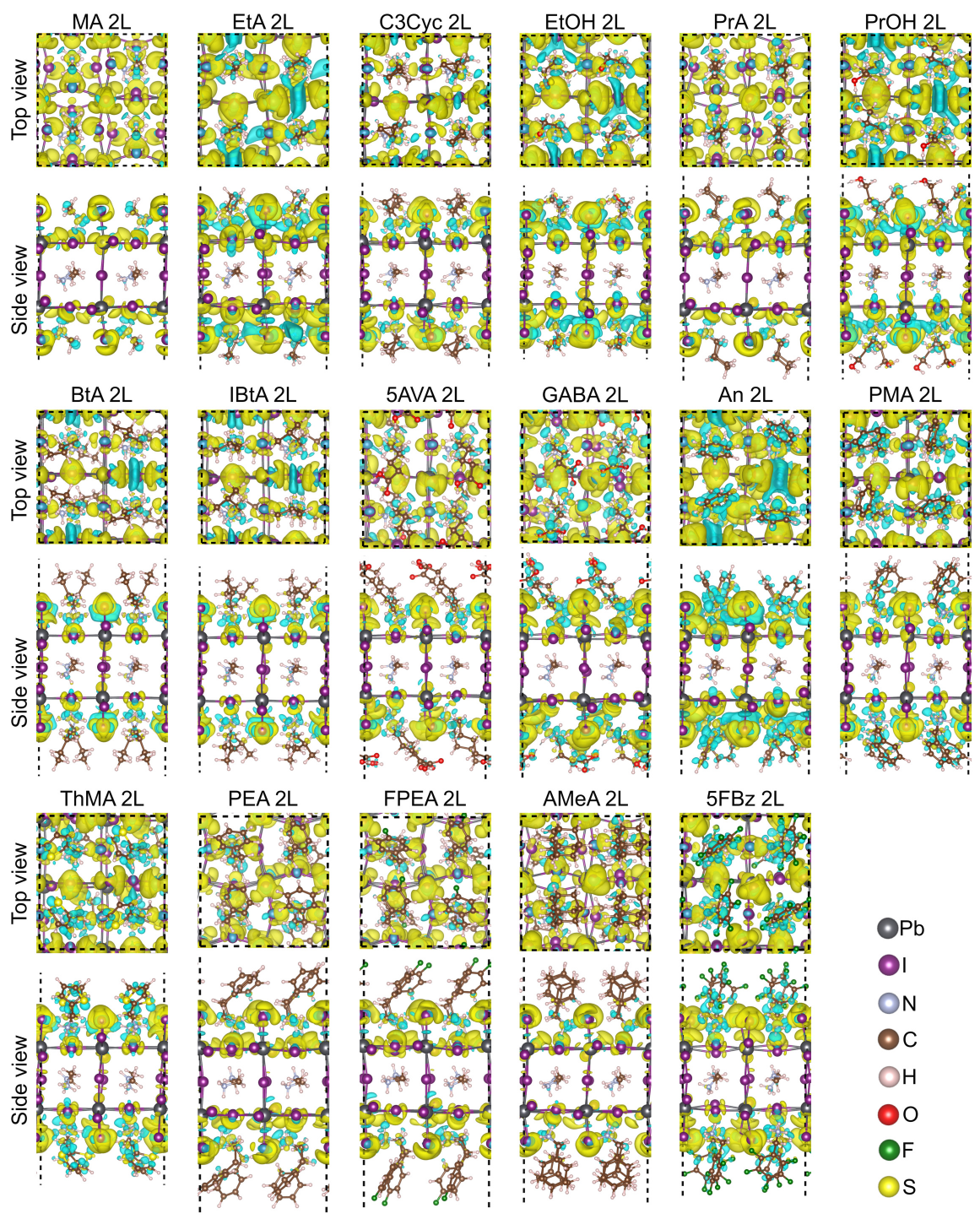}
    \caption{Electron density difference isosurfaces (\num{0.0015} $\si{bohr^{-3}}$) for bilayer perovskites. The yellow and blue regions indicate the accumulation and depletion of charge, respectively.}
    \label{fig:bi_EDD}
\end{figure*}

To improve our understanding on the charge transfer process, we performed an electron density difference analysis using the following equation,
\begin{equation}  
\Delta\rho = \rho^{\text{$P$~cation/slab}} - (\rho^{\text{$P$~cation}} + \rho^{\text{slab}})~,
\end{equation} 
where $\rho^{\text{$P$~cation/slab}}$ is the total electron density of 2D perovskites, $\rho^{\text{$P$~cation}} $ is the electron density of the isolated $P$-cation, $\rho^{\text{slab}}$ is the electron density of the inorganic slab, all fragments are frozen with respect to the optimized structure with equilibrium volume ($V_0$). For bilayers, the results $\Delta\rho$ are shown in Figure{~}\ref{fig:bi_EDD}, while the results for the remaining systems are shown in the supporting information.

The electron density difference results support the DDEC charge analyses, allowing us to present a comprehensive depiction of the charge-transfer mechanism. This involves a detailed examination of the accumulation and depletion of charges in specific sites in the 2D structures. In general, the values of $\Delta\rho$ suggest that there is an accumulation of charge in the halide. Furthermore, we observe different proportions of charge depletion along the $P$-cations for both monolayers and bilayers after thin film passivation, which confirms the transfer of charge from $P$-cations to iodides.

Due to the high electronegative character of fluoride, such systems exhibit substantial charge depletions near the ring carbons when we have a perfluorobenzen, 5FBz. This effect decreases when only a monosubstitution is considered in the \textit{para}-position, as in FPEA, and also has the effect of increasing the chain, influencing these results for this $P$-cation.

Similarly, charge depletions in carbons bonded to oxygenated groups, \ce{\bond{-}OH} and \ce{\bond{-}COOH}, support the assumption that such groups reduce the van der Waals gap and cause charge localization due to polarity and intermolecular interactions. To corroborate the notion of charge transfer through \ce{H+} transfer, we observe charge depletions in one of the \ce{H} bonds to \ce{N} in both organic molecules, $P$-cations, and MA interlayers. Furthermore, the observation that stabilization of organic cations occurs through interaction with the inorganic slab, \ce{N\bond{-}H\bond{...}I\bond{-}Pb}, is consistent with experimental data.\cite{Li_e202217910_2023}

In fact, we found charge accumulations near the \ce{N} atoms, highlighting the essential characteristic of the \ce{R\bond{-}NH3} group. These charge accumulations correlate with the values described above $Q^{\ce{R}}_{eff}$. In bilayer systems, MA interlayer molecules exhibit behavior similar to that observed in $P$-cations. This behavior results in charge depletion in one of the \ce{H} atoms of \ce{\bond{-}NH3} as well as in \ce{C} due to the electronegativity difference between \ce{C\bond{-}N}. Consequently, there is also charge accumulation near the \ce{N} atom for these molecules.

\subsection{Local Density of States}

\begin{figure*}[t!]
    \centering
    \includegraphics[width=0.80\linewidth]{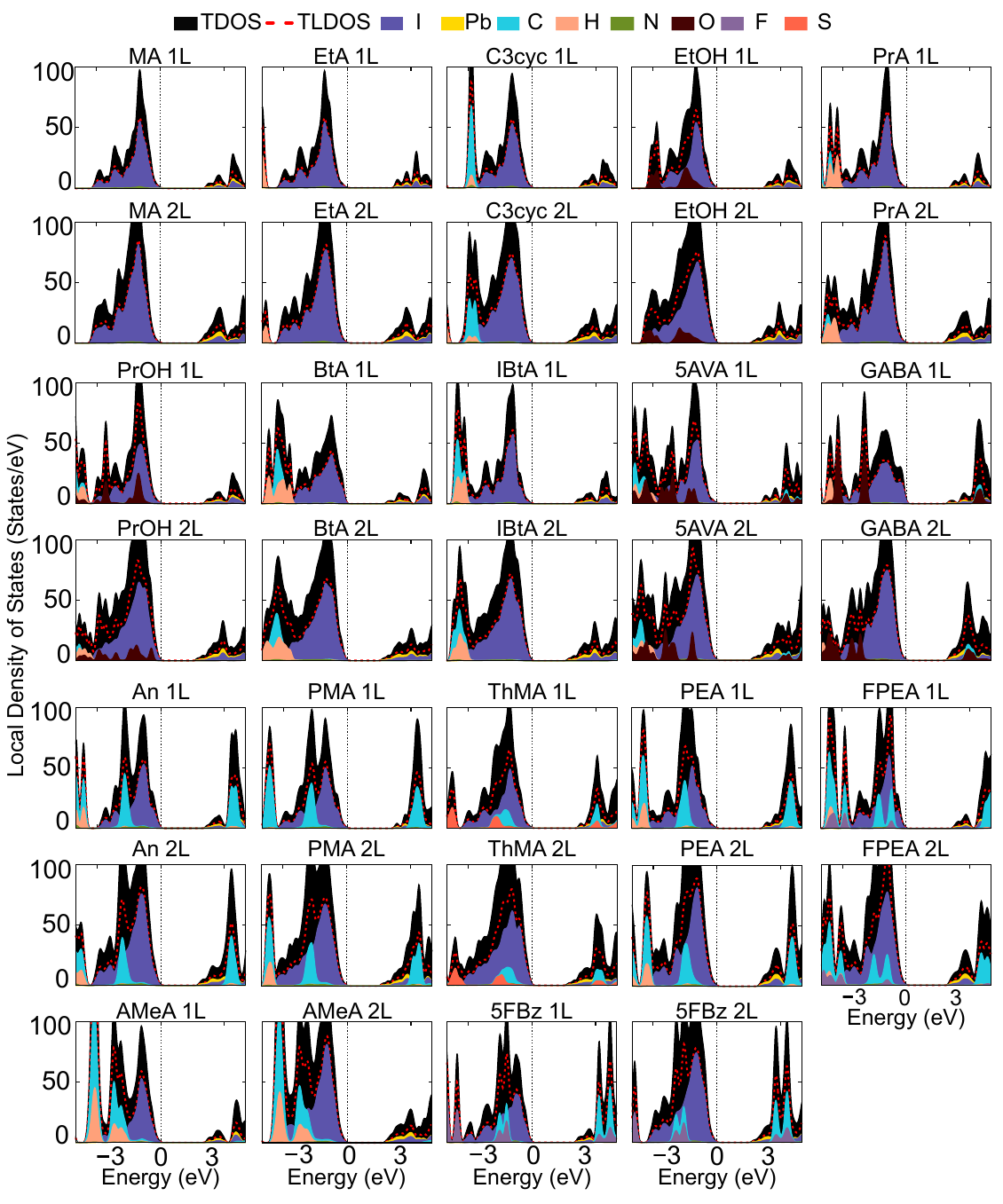}
    \caption{Local density of states for 2D perovskites-based materials, the valence band maximum was set at \SI{0}{\electronvolt} (vertical dashed line).}
    \label{fig:LDOS_2DPVK}
\end{figure*}

To characterize the electronic states near the band gap region (valence-band maximum (VBM) and conduction-band minimum (CBM) states), we conducted an in-depth analysis of the local density of states (LDOS). All results are shown in Figure{~}\ref{fig:LDOS_2DPVK}.

According to our findings, the band gap region for all structures has direct contributions from the $p$-states of iodide ions and the $s$- and $p$-states of lead atoms. Therefore, distortions of the \ce{PbI6}-octahedra play a crucial role in the energy band gap window of 2D perovskites. Thus, all of the descriptors that influence the distortions of the \ce{PbI6}-octahedra reported here affect the gap values of the structures.

We can also find contributions of electronic states to $P$-cations close to the band gap region for those containing cyclic aryl chains and electronegative groups (\ce{\bond{-}OH}, \ce{\bond{-}COOH}). According to LDOS, the contributions of the $p$-states of \ce{O} are shown in the VBM and CBM, and interactions through \ce{H}-bonds can influence the peak intensity and the energetic position of the states. In addition, we have contributions from the $p$-states to the structures with \ce{S} (ThMA) and \ce{F} (FPEA and 5FBz) atoms, which may be associated with the energetic stability of the system.\cite{Hu_171_2019}

The LDOS results for structures passivated with alkyl-type chain $P$-cations suggest the necessity of selecting cations with larger molecular radii. These structures exhibit contributions from the $p$-states of \ce{C} as a function of the size of the $P$-cation. It is possible to observe the appearance of the $p-$ and $s-$states of \ce{C} and \ce{H}, respectively, for $P$-cations composed of carbon chains with more than three carbons. The data also indicate that the $p$-states of \ce{N} are slightly above the states of \ce{C} and \ce{H} and slightly below the $p$-states of iodine. However, because of their low abundance compared to the total number of atoms in the system, the \ce{N} states exhibit low intensity.

From an electronic point of view, aromatic $P$-cations may outperform alkyl cations due to the symmetry properties of the $\pi$-bonds. In these systems, we found contributions from the organic molecule states closest to the VBM. However, we observed $p$-states of \ce{C} predominantly in the CBM. These findings imply that aromatic $P$-cations positively contribute to improved system stability and improve the charge transfer mechanism, thus improving hole mobility between organic and inorganic layers and, finally, increasing the conductivity of the 2D perovskite.\cite{Qin_1904050_2020}  

\subsection{Fundamental Energy Band Gaps}

\begin{figure}[t!]
    \centering
    \includegraphics[width=0.80\linewidth]{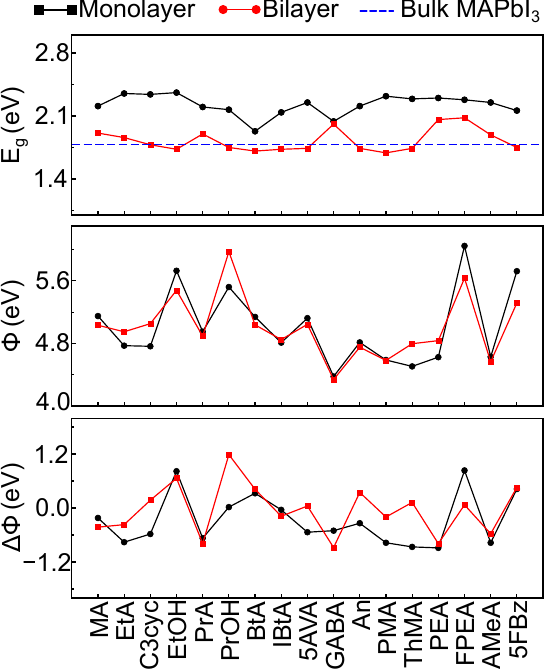}
    \caption{The PBE+D3 electronic band gap ($E_g$) for the 2D perovskites, in addition, we report as reference the \ce{MAPbI3} bulk band gap (dashed blue line) also obtained via PBE+D3 at $\Gamma$-point. In the mid panel, we have the work function ($\Phi$) and in the bottom panel the induced work function change ($\Delta \Phi$) both also calculated by the PBE+D3 level for all studied materials.}
    \label{fig:egap_wf}
\end{figure}

The electronic energy band gap ($E_g$) was measured by the energy difference between the VBM and the CBM. These data are important to characterize the photovoltaic performance of PSCs.\cite{Leng_2041_2020} Therefore, we qualitatively investigated the role of specific $P$-cations in the $E_g$ behavior. The corresponding results for $E_g$ are presented in Figure{~}\ref{fig:egap_wf}.

Our findings suggest that the second layer plays a role in the decrease in the values of $E_g$, as the quantum confinement effect is less pronounced in these systems compared to the corresponding monolayers.\cite{Cheng_1_2018} Therefore, the contribution of the states \ce{PbI6}-octahedra plays an important role in this reduction, which aligns with the findings of the LDOS analyzes.

However, variations in $E_g$ may result from intermolecular interactions. The GABA $P$-cation binding mechanism, similar in both systems, may be related to the slight differences in $E_g$ that we found for the monolayer and bilayer GABA systems. Furthermore, the presence of \ce{\bond{-}OH} and \ce{\bond{-}COOH} groups in molecular structures can induce a decrease in $E_g$ of 2D perovskites, compared to $P$-cations without oxygenated groups and the same number of carbons. Interestingly, synthesized crystals \ce{BtA2MAPb2I7} and \ce{GABA2MAPb2I7} crystals exhibit this similar pattern.\cite{Li_e202217910_2023}

\begin{figure*}[t!]
    \centering
    \includegraphics[width=0.95\linewidth]{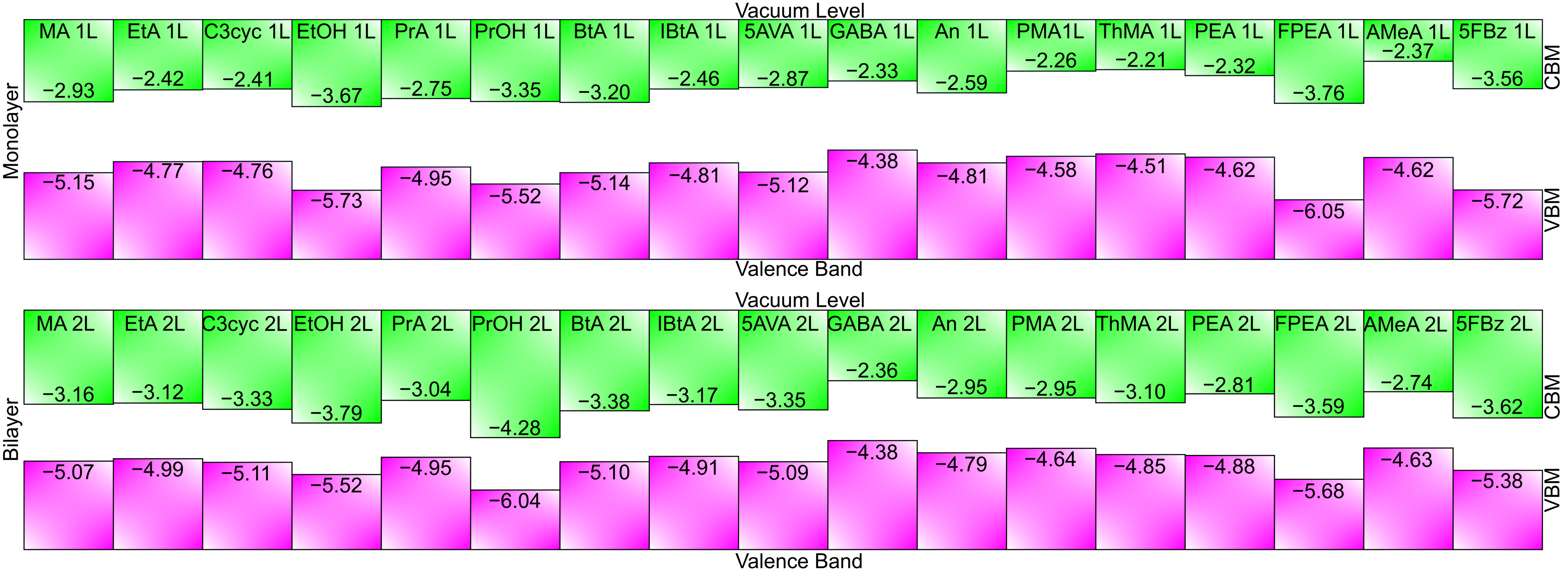}
    \caption{PBE+D3 VBM and CBM, in pink and green respectively, of 2D perovskites-based materials in equilibrium lattice parameters with respect to vacuum level (zero energy).}
    \label{fig:2DPVK_pbe_d3_band_offset}
\end{figure*}

\subsection{Work Function} 

The work function ($\Phi$) is a key parameter that characterizes the energy barrier for electron transfer between the perovskite and the electron transport layer in the PSCs.\cite{Leng_2041_2020} In this work, $\Phi$ is defined as the energy difference between the highest occupied state (VBM) and the electrostatic potential value within the vacuum region. We also evaluate the induced change in work function with respect to the clean surface slab, $\Delta\Phi = \Phi^{\text{$P$~cation/slab}} - \Phi^{\text{slab}}$. All the results of $\Phi$ and $\Delta\Phi$ are shown in Figure{~}\ref{fig:egap_wf}. 

Our findings reveal that in general $\Phi$ is highly sensitive to surface conditions, depending exclusively on the type of $P$-cation adsorbed in the system, making $\Phi$ thickness independent. $P$-cations with terminal atoms more electronegative than iodides (e.g. oxygen and fluorine) increase the value of $\Phi$ due to the influence on charge transfer and dipole moment. In this way, we observe an increase in $\Phi$ of \SI{9.67}{\percent}, \SI{18.09}{\percent}  and \SI{14.16}{\percent}, respectively, for the EtOH, PrOH, and FPEA bilayers compared to their respective $P$-cation bilayers without electronegative atoms (EtA, PrA, and PEA). Consequently, there is a reduction in $\Phi$ for the terminal atoms of ligands that exhibit higher electropositivity. These findings corroborate the existing literature; Zheng \textit{et al.} demonstrated a decrease in $\Phi$ as $X$ is replaced in the order of \ce{F}, \ce{Cl}, \ce{Br} and \ce{I} within the molecular structure of 3-$X$PEA when used as the $P$-cation.\cite{Zheng_2328_2021} 

However, there exists a difference in effects when comparing $P$-cations with terminal groups \ce{\bond{-}OH} and \ce{\bond{-}COOH}, with an increase in $\Phi$ observed for $P$-cations possessing \ce{\bond{-}OH} and a reduction in $\Phi$ for those bearing \ce{\bond{-}COOH}, as demonstrated by ${\Delta}\Phi$. This can be elucidated by the different acidic nature of $P$-cations with \ce{\bond{-}OH} and \ce{\bond{-}COOH} groups, the latter exhibiting a greater acidic character and, as a result, the ability to dissociate easily to release a hydrogen ion and therefore decreasing the $\Phi$. Therefore, the charge transfer mechanism significantly influences the results of $\Phi$.

\subsection{Band off Set} 

The band offset for all 2D perovskites was achieved by adjusting their eigenvalues in relation to the vacuum energy level. The results are shown in Figure{~}\ref{fig:2DPVK_pbe_d3_band_offset}. The band alignment of 2D perovskites is influenced by two distinct phenomena: $(i)$ the passivation effects of $P$ cations and $(ii)$ the quantum confinement effects. In this way, we investigated these effects and their implications for the electronic structure and properties of these materials.

According to the LDOS data, the chemistry of $P$-cations has a significant impact on band alignment because of its contributions from states near the band gap region, which affects the location of valence states in thin films. The VBM and CBM shifts towards lower energy levels are visible in $P$-cations functionalized with OH groups. The behavior of the 2L system mirrors that of the 1L EtOH system, with a VBM and CBM energy drop of \SI{16.71}{\percent} and \SI{28.12}{\percent}, respectively, compared to the EtA system. This impact is also evident when the $P$-cation is functionalized with fluorine (FPEA), resulting in a \SI{23.52}{\percent} decrease in VBM and a \SI{38.29}{\percent} decrease in CBM compared to the system with PEA as the $P$-cation, for the same thickness.

Comparing the 1L and 2L systems reveals a higher percentage variation in the CBM values; as a result, the influence of quantum confinement is more perceptible in the CBM regions, changing this eigenvalue to more negative values and reducing the band gap values in 2L systems. The energy states created by quantum confinement can facilitate charge separation and transport within 2D perovskite heterostructures. When the layer thickness is tuned, it becomes possible to design heterostructures with favorable band alignments for efficient charge carrier dynamics, offering promise for applications such as PSCs.

\subsection{Absorption Coefficient and Optical Band Gap} 

\begin{figure}[t!]
    \centering
    \includegraphics[width=0.90\linewidth]{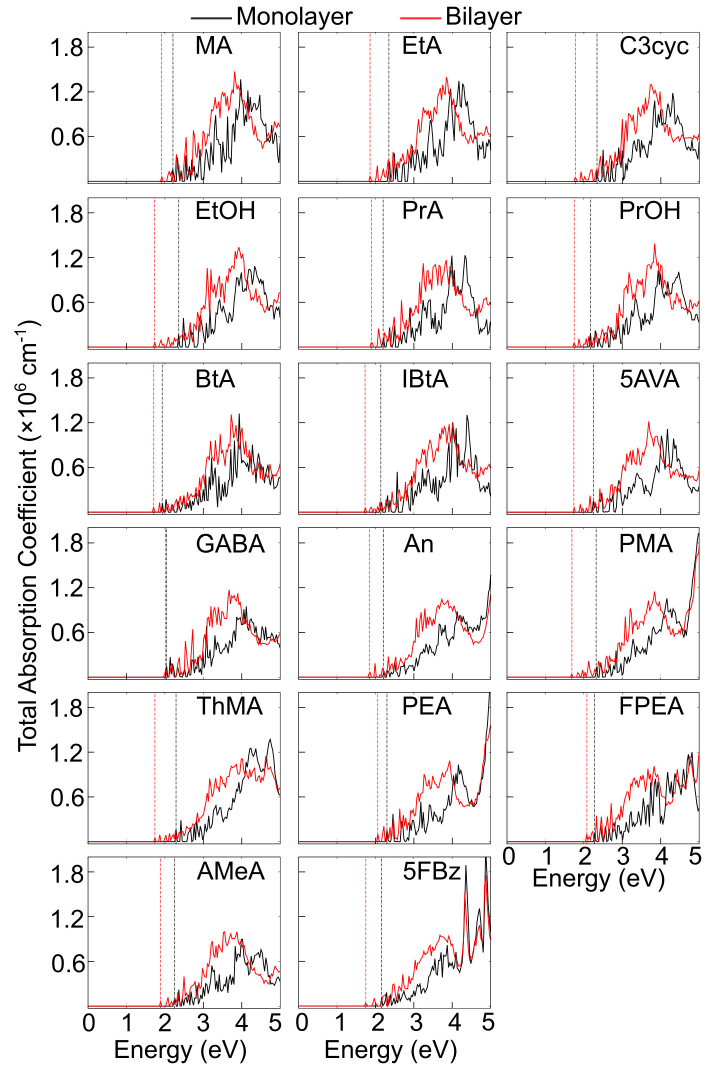}
    \caption{Optical absorption coefficient, obtained from PBE+D3, of the monolayer (black line) and bilayer (red line) 2D perovskites. $E_{g}^{optic}$ values are in indicated by dashed lines}
    \label{fig:abs_coefficient}
\end{figure}

The absorption spectrum is qualitatively investigated using the PBE+D3 level. All results are shown in Figure{~}\ref{fig:abs_coefficient}. The behavior of the optical absorption coefficient spectra here responds as a function of the passivation of the selected $P$-cation, similar to the electronic descriptors studied previously. Furthermore, we observe that the absorption spectra exhibit a trend similar to that of $E_{g}$, with the emission wavelength increasing as the number of $n$-layers rises.

The optical band gap $(E_{g}^{optic})$ can be defined by the first allowed electronic transition with nonzero magnitude.\cite{dosSantos_5259_2023} We found that $E_{g}^{optic}$ is equivalent to the fundamental band gaps $(E_{g})$ obtained from the predicted absorption coefficient spectra. These spectra only include direct transitions from the valence band to the conduction band. This observation supports the idea that the promotion of electrons from the VBM to the CBM through energy photon absorption occurs for our models. It should be noted that our models do not have structural defects (such as point defects, edges, and grain boundaries), which are typically considered in experimental absorption spectra.\cite{Liu_3335_2016} Thus, surface passivation with an organic ammonium cation is effective in modifying the structural and optoelectronic properties of materials based on 2D perovskites, leading to improved device performance.

\section{Insights into the Selection of $P$-Cations for the Passivation of Two-Dimensional Perovskites}

In summary, our calculations have provided a fundamental characterization of the structural and optoelectronic properties of 2D thin films perovskites against different types of passivating organic molecules. Therefore, the choice of the best $P$-cation naturally starts with the main application of the material. However, on the basis of our results, we can give some general directions for the choice of $P$-cation for optoelectronic applications.

\paragraph{Alkylammonium $P$-cations molecules:} Passivation by aliphatic organic cations with short chains, such as \ce{(NR_{x}H_{4-x})+} with $x = 1-2$, may have drawbacks compared to organic compounds with longer chains. Our findings indicate that the energy stability of perovskite-based materials increases with increasing carbon chain of alkylammonium $P$-cations. Ye \textit{et al.},\cite{Ye_1566_2022} show that passivation using $P$-cations with \num{5} carbons in the primary organic chain improves the PCE of PSCs compared to data for $P$-cations with \num{4} and \num{2} carbons. Importantly, $P$-cations from short carbon chains exhibit instabilities when exposed to humidity and temperature conditions, which can cause structural defects and reduce the efficiency of the material.

\paragraph{Alcohol and Carboxylic acids $P$-cations molecules:} Organic cations functionalized with \ce{\bond{-}OH} and \ce{\bond{-}COOH} groups promote additional intermolecular interactions compared to molecular structures without these functional groups, achieved through \ce{H}-bonds. Our results demonstrate an increase in energetic stability via passivation by such $P$-cations. Furthermore, the inclusion of oxygenated groups promotes \ce{H}-bond interactions while reducing the van der Waals gap, leading to improved optoelectronic characteristics. Charges are effectively located within the $P$-cation sites, enabling the formation of "charged bridges", as also noted by Li \textit{et al.}\cite{Li_e202217910_2023} This effect, in turn, enhances charge concentrations, promoting overlap in adjacent inorganic layers and, thus, optimizing charge transfer efficiency between them.

\paragraph{Aromatic $P$-cations molecules:} Aromatic molecules can offer distinct benefits when compared to aliphatic cations due to their bulkier conjugated structure.\cite{Gao_1_2023} The conjugated structure of aromatic cations enhances the stability of 2D perovskites and controls crystal arrangement and orientation via \ce{\pi\bond{...}\pi} interactions and \ce{H} bonds. This results in a more stable 2D perovskite by minimizing the slippage of inorganic slabs along the van der Waals gap direction.

\paragraph{Fluorinated $P$-cations molecules:} According to our findings, $P$-cations with high number of \ce{F} atoms in their molecular structure can reduce energy stability via $E_{ad}$ and $E_{int}$ energies, as observed in the 5FBz system containing \num{5} \ce{F} atoms bound to the aromatic ring, resulting in interhalide repulsion energies. However, as the \ce{F} content in the structure decreases, this impact decreases linearly, as seen in the FPEA system. The \ce{F} bound at the \textit{para} position to the aromatic ring can induce structural stability through dipole-dipole interactions, encouraging stronger \ce{H}-bonds than $\ce{H\bond{...}I}$-type interactions. In a study by Liu \textit{et al.},\cite{Li_2072_2021} it was shown that the 2D perovskite with the aromatic formamidinium fluorinated at the \textit{para} position, namely p-FPhFA, exhibits significantly longer charge carrier lifetime, hole mobility, and lower trap density compared to a 2D perovskite with an unfluorinated $P$-cation.

Based on the insights presented, we selected three main $P$-cations with the greatest potential for application in PSC: FPEA, PEA, and BtA. We believe that such molecular structures can improve the properties of thin films during passivation, thus resulting in 2D structures with greater performance.

\section{Conclusions}

We performed DFT calculations to investigate the role of organic monocations on the passivation of thin films of 2D perovskites. The perovskites have a general chemical formula of \ce{$P$_{2}($MA$_{$n$–1})Pb_{$n$}I_{$3n+1$}}, where $n = 1$ (monolayer) and $n = 2$ (bilayer). Our findings suggest that passivation has a significant impact on the structural and optoelectronic properties of ultrathin perovskites films, which can provide insight into the performance of the materials.

Interactions between organic cations and the anionic surface of 2D perovskite films promote distortions in the lattice parameters, which are due to the physicochemical character of the organic molecule acting as $P$-cation. For example, the 5FBz bilayer system showed the lowest  ECN$_{av}^{\ce{Pb}}$ value, approximately \SI{4.17}{\percent} smaller than the bulk, probably due to distortions in the center cation displacement caused by highly electronegative fluorine ions in the $P$-cation. Our results indicate that the size of the $P$-cation and the presence of intermolecular and intramolecular interactions influence the lattice parameters, which can provide insight into the energetic stability of compounds, including \ce{H}-bonds. Therefore, different $P$-cations induce different deformations in the distances and bond angles within the \ce{PbI6}-octahedra. 

Our study delves into the dynamics of charge transfer from the $P$-cations to the inorganic framework of the systems. The binding mechanism results in the charge transfer from the \ce{\bond{-}NH3} group to the inorganic slab, causing charge accumulations in halides and indicating the behavior of the Br\o nsted--Lowry acid base between $P$-cations and slabs. Consequently, charge depletions are observed near the sites of \ce{H} and \ce{C} bonded to electronegative groups. In particular, our results demonstrated that the aromatic $P$-cation cations showed the impact of electronic characteristics of the $\pi$-bond resonance, for example, we found an increase of approximately \SI{5}{\percent} in $Q^{P}_{eff}$ when comparing the effects for the \ce{An2PbI4} and \ce{PEA2PbI4} films. Radicals with highly electronegative groups induce localized interlayer charges, which contributes to higher $Q^{P}_{eff}$ values. Quantum confinement effects in 2D systems affect the charge properties of inorganic slabs, with lower $Q^{Pb}_{eff}$ values, suggesting that increasing the number of inorganic layers can enhance the negative charge on the halides, indicating the photovoltaic potential. 

Our findings show that the band gap region in all structures is directly influenced by the $p$-states of iodides and the $s$- and $p$-states of lead. Structural descriptors that affect \ce{PbI6}-octahedra distortions have a direct impact on gap values. Further analysis reveals significant electronic state contributions from $P$-cations near the band gap region, especially for cyclic aryl chains and electronegative groups (\ce{\bond{-}OH}, \ce{\bond{-}COOH}). LDOS indicates prominent \ce{O} $p$-states in both the VBM and the CBM, influenced by \ce{H}-bond interactions. Additional contributions from the \ce{S} (ThMA) and \ce{F} (FPEA and 5FBz) atoms are observed, potentially contributing to energetic stability. The LDOS results advocate for larger molecular radii in cations, especially for alkyl-type chain $P$-cations. Electronic analysis suggests that aromatic $P$-cations outperform alkyl cations due to the symmetry properties of $\pi$-bonds, enhancing the stability of the system, charge transfer, hole mobility and conductivity in 2D perovskites films. 

From the results, we selected three main $P$-cations for PSC application: FPEA, PEA, and BtA. These organic structures can improve thin film properties during passivation, enhancing perovskites-based devices' performance.

\begin{acknowledgement}
The authors acknowledge Dr. Srikanth Malladi for his contributions to the construction of the initial models, which played a crucial role for the success of the present study. The authors express their gratitude for the support received from various organizations. They appreciate the assistance provided by FAPESP (S{\~a}o Paulo Research Foundation, Brazil) through grant numbers $2017/11631$-$2$ and $2018/21401$-$7$, Shell and the strategic importance of the support provided by ANP (Brazil's National Oil, Natural Gas and Biofuels Agency) through the R\&D levy regulation. The ICR extends its thanks to the National Council for Scientific and Technological Development (CNPq) for the Ph.D. fellowship, which was granted under the numbers $140015/2021$-$3$. AFAB also acknowledges financial support from FAPESP, grant numbers $2017/11937$-$4$ and $2022/12778$-$5$ (Postdoctoral fellowship) and PIRM acknowledges financial support from FAPESP, grant number $2023/12824$-$0$ (Postdoctoral fellowship). The authors are also grateful for the infrastructure support offered by the Department of Information Technology - Campus S{\~a}o Carlos for our computer cluster.
\end{acknowledgement}

\begin{suppinfo}
The data used for the figures, as well as additional analyzes and technical details, are reported in the electronic supplementary material. 
\end{suppinfo}

\bibliography{israelribeiro-pedro_Mol-2DPVK.bbl}

\providecommand{\latin}[1]{#1}
\makeatletter
\providecommand{\doi}
  {\begingroup\let\do\@makeother\dospecials
  \catcode`\{=1 \catcode`\}=2 \doi@aux}
\providecommand{\doi@aux}[1]{\endgroup\texttt{#1}}
\makeatother
\providecommand*\mcitethebibliography{\thebibliography}
\csname @ifundefined\endcsname{endmcitethebibliography}
  {\let\endmcitethebibliography\endthebibliography}{}
\begin{mcitethebibliography}{69}
\providecommand*\natexlab[1]{#1}
\providecommand*\mciteSetBstSublistMode[1]{}
\providecommand*\mciteSetBstMaxWidthForm[2]{}
\providecommand*\mciteBstWouldAddEndPuncttrue
  {\def\EndOfBibitem{\unskip.}}
\providecommand*\mciteBstWouldAddEndPunctfalse
  {\let\EndOfBibitem\relax}
\providecommand*\mciteSetBstMidEndSepPunct[3]{}
\providecommand*\mciteSetBstSublistLabelBeginEnd[3]{}
\providecommand*\EndOfBibitem{}
\mciteSetBstSublistMode{f}
\mciteSetBstMaxWidthForm{subitem}{(\alph{mcitesubitemcount})}
\mciteSetBstSublistLabelBeginEnd
  {\mcitemaxwidthsubitemform\space}
  {\relax}
  {\relax}

\bibitem[Kojima \latin{et~al.}(2009)Kojima, Teshima, Shirai, and
  Miyasaka]{Kojima_6050_2009}
Kojima,~A.; Teshima,~K.; Shirai,~Y.; Miyasaka,~T. {Organometal Halide
  Perovskites as Visible-Light Sensitizers for Photovoltaic Cells}. \emph{J.
  Am. Chem. Soc.} \textbf{2009}, \emph{131}, 6050--6051, DOI:
  \doi{10.1021/ja809598r}\relax
\mciteBstWouldAddEndPuncttrue
\mciteSetBstMidEndSepPunct{\mcitedefaultmidpunct}
{\mcitedefaultendpunct}{\mcitedefaultseppunct}\relax
\EndOfBibitem
\bibitem[Gao \latin{et~al.}(2023)Gao, Li, Li, Wu, Zhang, Zhao, Jiang, Zhang,
  Wang, Li, Li, Xiao, Choy, Jen, Yang, and Zhu]{Gao_2206387_2023}
Gao,~D.; Li,~B.; Li,~Z.; Wu,~X.; Zhang,~S.; Zhao,~D.; Jiang,~X.; Zhang,~C.;
  Wang,~Y.; Li,~Z.; Li,~N.; Xiao,~S.; Choy,~W. C.~H.; Jen,~A. K.-Y.; Yang,~S.;
  Zhu,~Z. {Highly Efficient Flexible Perovskite Solar Cells through
  Pentylammonium Acetate Modification with Certified Efficiency of 23.35{\%}}.
  \emph{Adv. Mater.} \textbf{2023}, \emph{35}, 2206387, DOI:
  \doi{10.1002/adma.202206387}\relax
\mciteBstWouldAddEndPuncttrue
\mciteSetBstMidEndSepPunct{\mcitedefaultmidpunct}
{\mcitedefaultendpunct}{\mcitedefaultseppunct}\relax
\EndOfBibitem
\bibitem[Zheng \latin{et~al.}(2024)Zheng, Li, Zhuang, Wu, Tian, Sun, Chen, Guo,
  Hua, Meng, Wu, and Chen]{Zheng_1153_2024}
Zheng,~Y.; Li,~Y.; Zhuang,~R.; Wu,~X.; Tian,~C.; Sun,~A.; Chen,~C.; Guo,~Y.;
  Hua,~Y.; Meng,~K.; Wu,~K.; Chen,~C.-C. {Towards 26{\%} Efficiency in Inverted
  Perovskite Solar Cells via Interfacial Flipped Band Bending and Suppressed
  Deep-level Traps}. \emph{Energy Environ. Sci.} \textbf{2024}, \emph{17},
  1153--1162, DOI: \doi{10.1039/D3EE03435F}\relax
\mciteBstWouldAddEndPuncttrue
\mciteSetBstMidEndSepPunct{\mcitedefaultmidpunct}
{\mcitedefaultendpunct}{\mcitedefaultseppunct}\relax
\EndOfBibitem
\bibitem[De~Wolf \latin{et~al.}(2014)De~Wolf, Holovsky, Moon,
  L{\ifmmode\ddot{o}\else\"{o}\fi}per, Niesen, Ledinsky, Haug, Yum, and
  Ballif]{DeWolf_1035_2014}
De~Wolf,~S.; Holovsky,~J.; Moon,~S.-J.;
  L{\ifmmode\ddot{o}\else\"{o}\fi}per,~P.; Niesen,~B.; Ledinsky,~M.;
  Haug,~F.-J.; Yum,~J.-H.; Ballif,~C. {Organometallic Halide Perovskites: Sharp
  Optical Absorption Edge and Its Relation to Photovoltaic Performance}.
  \emph{J. Phys. Chem. Lett.} \textbf{2014}, \emph{5}, 1035--1039, DOI:
  \doi{10.1021/jz500279b}\relax
\mciteBstWouldAddEndPuncttrue
\mciteSetBstMidEndSepPunct{\mcitedefaultmidpunct}
{\mcitedefaultendpunct}{\mcitedefaultseppunct}\relax
\EndOfBibitem
\bibitem[Park(2015)]{Park_65_2015}
Park,~N.-G. Perovskite Solar Cells: An Emerging Photovoltaic Technology.
  \emph{Mater. Today} \textbf{2015}, \emph{18}, 65--72, DOI:
  \doi{10.1016/j.mattod.2014.07.007}\relax
\mciteBstWouldAddEndPuncttrue
\mciteSetBstMidEndSepPunct{\mcitedefaultmidpunct}
{\mcitedefaultendpunct}{\mcitedefaultseppunct}\relax
\EndOfBibitem
\bibitem[Zhang \latin{et~al.}(2013)Zhang, Saliba, Stranks, Sun, Shi, Wiesner,
  and Snaith]{Zhang_4505_2013}
Zhang,~W.; Saliba,~M.; Stranks,~S.~D.; Sun,~Y.; Shi,~X.; Wiesner,~U.;
  Snaith,~H.~J. {Enhancement of Perovskite-Based Solar Cells Employing
  Core{\textendash}Shell Metal Nanoparticles}. \emph{Nano Lett.} \textbf{2013},
  \emph{13}, 4505--4510, DOI: \doi{10.1021/nl4024287}\relax
\mciteBstWouldAddEndPuncttrue
\mciteSetBstMidEndSepPunct{\mcitedefaultmidpunct}
{\mcitedefaultendpunct}{\mcitedefaultseppunct}\relax
\EndOfBibitem
\bibitem[Ponseca \latin{et~al.}(2014)Ponseca, Savenije, Abdellah, Zheng,
  Yartsev, Pascher, Harlang, Chabera, Pullerits, Stepanov, Wolf, and
  Sundstr{\ifmmode\ddot{o}\else\"{o}\fi}m]{Ponseca_5189_2014}
Ponseca,~C.~S.,~Jr.; Savenije,~T.~J.; Abdellah,~M.; Zheng,~K.; Yartsev,~A.;
  Pascher,~T.; Harlang,~T.; Chabera,~P.; Pullerits,~T.; Stepanov,~A.;
  Wolf,~J.-P.; Sundstr{\ifmmode\ddot{o}\else\"{o}\fi}m,~V. {Organometal Halide
  Perovskite Solar Cell Materials Rationalized: Ultrafast Charge Generation,
  High and Microsecond-Long Balanced Mobilities, and Slow Recombination}.
  \emph{J. Am. Chem. Soc.} \textbf{2014}, \emph{136}, 5189--5192, DOI:
  \doi{10.1021/ja412583t}\relax
\mciteBstWouldAddEndPuncttrue
\mciteSetBstMidEndSepPunct{\mcitedefaultmidpunct}
{\mcitedefaultendpunct}{\mcitedefaultseppunct}\relax
\EndOfBibitem
\bibitem[Stoumpos \latin{et~al.}(2013)Stoumpos, Malliakas, and
  Kanatzidis]{Stoumpos_9019_2013}
Stoumpos,~C.~C.; Malliakas,~C.~D.; Kanatzidis,~M.~G. Semiconducting Tin and
  Lead Iodide Perovskites With Organic Cations: Phase Transitions, High
  Mobilities, and Near-Infrared Photoluminescent Properties. \emph{Inorg.
  Chem.} \textbf{2013}, \emph{52}, 9019--9038, DOI:
  \doi{10.1021/ic401215x}\relax
\mciteBstWouldAddEndPuncttrue
\mciteSetBstMidEndSepPunct{\mcitedefaultmidpunct}
{\mcitedefaultendpunct}{\mcitedefaultseppunct}\relax
\EndOfBibitem
\bibitem[Lekina and Shen(2019)Lekina, and Shen]{Lekina_189_2019}
Lekina,~Y.; Shen,~Z.~X. Excitonic States and Structural Stability in
  Two-Dimensional Hybrid Organic-Inorganic Perovskites. \emph{J. Sci.: Adv.
  Mater. Dev.} \textbf{2019}, \emph{4}, 189--200, DOI:
  \doi{10.1016/j.jsamd.2019.03.005}\relax
\mciteBstWouldAddEndPuncttrue
\mciteSetBstMidEndSepPunct{\mcitedefaultmidpunct}
{\mcitedefaultendpunct}{\mcitedefaultseppunct}\relax
\EndOfBibitem
\bibitem[Paritmongkol \latin{et~al.}(2019)Paritmongkol, Dahod, Stollmann, Mao,
  Settens, Zheng, and Tisdale]{Paritmongkol_5592_2019}
Paritmongkol,~W.; Dahod,~N.~S.; Stollmann,~A.; Mao,~N.; Settens,~C.;
  Zheng,~S.-L.; Tisdale,~W.~A. Synthetic Variation and Structural Trends in
  Layered Two-dimensional Alkylammonium Lead Halide Perovskites. \emph{Chem.
  Mater.} \textbf{2019}, \emph{31}, 5592--5607, DOI:
  \doi{10.1021/acs.chemmater.9b01318}\relax
\mciteBstWouldAddEndPuncttrue
\mciteSetBstMidEndSepPunct{\mcitedefaultmidpunct}
{\mcitedefaultendpunct}{\mcitedefaultseppunct}\relax
\EndOfBibitem
\bibitem[Danelon \latin{et~al.}(2024)Danelon, Santos, Dias, Da~Silva, and
  Lima]{Danelon_8469_2024}
Danelon,~J.~G.; Santos,~R.~M.; Dias,~A.~C.; Da~Silva,~J. L.~F.; Lima,~M.~P.
  {Contrasting the Stability, Octahedral Distortions, and Optoelectronic
  Properties of 3D \ce{MABX3} and 2D \ce{(BA)2(MA)B2X7} (B = Ge, Sn, Pb; X =
  Cl, Br, I) Perovskites}. \emph{Phys. Chem. Chem. Phys.} \textbf{2024},
  \emph{26}, 8469--8487, DOI: \doi{10.1039/D3CP04361D}\relax
\mciteBstWouldAddEndPuncttrue
\mciteSetBstMidEndSepPunct{\mcitedefaultmidpunct}
{\mcitedefaultendpunct}{\mcitedefaultseppunct}\relax
\EndOfBibitem
\bibitem[Zhang \latin{et~al.}(2023)Zhang, Tai, Li, Zhao, Chen, Yin, Zhou,
  Zhang, Han, Wang, and Lin]{Zhang_2023}
Zhang,~M.; Tai,~M.; Li,~X.; Zhao,~X.; Chen,~H.; Yin,~X.; Zhou,~Y.; Zhang,~Q.;
  Han,~J.; Wang,~N.; Lin,~H. {Improved Moisture Stability of Perovskite Solar
  Cells Using N719 Dye Molecules}. \emph{Solar RRL} \textbf{2023}, \emph{3},
  DOI: \doi{10.1002/solr.201900345}\relax
\mciteBstWouldAddEndPuncttrue
\mciteSetBstMidEndSepPunct{\mcitedefaultmidpunct}
{\mcitedefaultendpunct}{\mcitedefaultseppunct}\relax
\EndOfBibitem
\bibitem[Sun \latin{et~al.}(2017)Sun, Fang, Ma, Xu, Lu, Yu, Yuan, and
  Ding]{Sun_8682_2017}
Sun,~Y.; Fang,~X.; Ma,~Z.; Xu,~L.; Lu,~Y.; Yu,~Q.; Yuan,~N.; Ding,~J. {Enhanced
  UV-Light Stability of Organometal Halide Perovskite Solar Cells with
  Interface Modification and a UV Absorption Layer}. \emph{J. Mater. Chem. C}
  \textbf{2017}, \emph{5}, 8682--8687, DOI: \doi{10.1039/C7TC02603J}\relax
\mciteBstWouldAddEndPuncttrue
\mciteSetBstMidEndSepPunct{\mcitedefaultmidpunct}
{\mcitedefaultendpunct}{\mcitedefaultseppunct}\relax
\EndOfBibitem
\bibitem[Chi and Banerjee(2021)Chi, and Banerjee]{Chi_0897_2021}
Chi,~W.; Banerjee,~S.~K. {Achieving Resistance against Moisture and Oxygen for
  Perovskite Solar Cells with High Efficiency and Stability}. \emph{Chem.
  Mater.} \textbf{2021}, \emph{33}, 4269--4303, DOI:
  \doi{10.1021/acs.chemmater.1c00773}\relax
\mciteBstWouldAddEndPuncttrue
\mciteSetBstMidEndSepPunct{\mcitedefaultmidpunct}
{\mcitedefaultendpunct}{\mcitedefaultseppunct}\relax
\EndOfBibitem
\bibitem[Adil~Afroz \latin{et~al.}(2020)Adil~Afroz, Ghimire, Reza, Bahrami,
  Bobba, Gurung, Chowdhury, Iyer, and Qiao]{AdilAfroz_2432_2020}
Adil~Afroz,~M.; Ghimire,~N.; Reza,~K.~M.; Bahrami,~B.; Bobba,~R.~S.;
  Gurung,~A.; Chowdhury,~A.~H.; Iyer,~P.~K.; Qiao,~Q. {Thermal Stability and
  Performance Enhancement of Perovskite Solar Cells Through Oxalic Acid-Induced
  Perovskite Formation}. \emph{ACS Appl. Energy Mater.} \textbf{2020},
  \emph{3}, 2432--2439, DOI: \doi{10.1021/acsaem.9b02111}\relax
\mciteBstWouldAddEndPuncttrue
\mciteSetBstMidEndSepPunct{\mcitedefaultmidpunct}
{\mcitedefaultendpunct}{\mcitedefaultseppunct}\relax
\EndOfBibitem
\bibitem[Wang \latin{et~al.}(2018)Wang, Cao, Chen, Chen, Wu, Li, Qin, and
  Huang]{Wang_1803753_2018}
Wang,~F.; Cao,~Y.; Chen,~C.; Chen,~Q.; Wu,~X.; Li,~X.; Qin,~T.; Huang,~W.
  Materials Toward the Upscaling of Perovskite Solar Cells: Progress,
  Challenges, and Strategies. \emph{Adv. Funct. Mater.} \textbf{2018},
  \emph{28}, 1803753, DOI: \doi{10.1002/adfm.201803753}\relax
\mciteBstWouldAddEndPuncttrue
\mciteSetBstMidEndSepPunct{\mcitedefaultmidpunct}
{\mcitedefaultendpunct}{\mcitedefaultseppunct}\relax
\EndOfBibitem
\bibitem[dos Santos \latin{et~al.}(2023)dos Santos, Ornelas{-}Cruz, Dias, Lima,
  and Da~Silva]{dosSantos_5259_2023}
dos Santos,~R.~M.; Ornelas{-}Cruz,~I.; Dias,~A.~C.; Lima,~M.~P.; Da~Silva,~J.
  L.~F. {Theoretical Investigation of the Role of Mixed \ce{A+} Cations in the
  Structure, Stability, and Electronic Properties of Perovskite Alloys}.
  \emph{ACS Appl. Energy Mater.} \textbf{2023}, \emph{6}, 5259--5273, DOI:
  \doi{10.1021/acsaem.3c00186}\relax
\mciteBstWouldAddEndPuncttrue
\mciteSetBstMidEndSepPunct{\mcitedefaultmidpunct}
{\mcitedefaultendpunct}{\mcitedefaultseppunct}\relax
\EndOfBibitem
\bibitem[Ribeiro \latin{et~al.}(2023)Ribeiro, Moraes, Bittencourt, and
  Da~Silva]{Ribeiro_13667_2023}
Ribeiro,~I.~C.; Moraes,~P. I.~R.; Bittencourt,~A. F.~B.; Da~Silva,~J. L.~F.
  {Role of the Adsorption of Alkali Cations on Ultrathin n-Layers of
  Two-Dimensional Perovskites}. \emph{J. Phys. Chem. C} \textbf{2023},
  \emph{127}, 13667--13677, DOI: \doi{10.1021/acs.jpcc.3c01894}\relax
\mciteBstWouldAddEndPuncttrue
\mciteSetBstMidEndSepPunct{\mcitedefaultmidpunct}
{\mcitedefaultendpunct}{\mcitedefaultseppunct}\relax
\EndOfBibitem
\bibitem[Hu \latin{et~al.}(2019)Hu, Oswald, Hu, Stuard, Nahid, Yan, Chen, Ade,
  Neilson, and You]{Hu_171_2019}
Hu,~J.; Oswald,~I. W.~H.; Hu,~H.; Stuard,~S.~J.; Nahid,~M.~M.; Yan,~L.;
  Chen,~Z.; Ade,~H.; Neilson,~J.~R.; You,~W. {Aryl-Perfluoroaryl Interaction in
  Two-Dimensional Organic{\textendash}Inorganic Hybrid Perovskites Boosts
  Stability and Photovoltaic Efficiency}. \emph{ACS Mater. Lett.}
  \textbf{2019}, \emph{1}, 171--176, DOI:
  \doi{10.1021/acsmaterialslett.9b00102}\relax
\mciteBstWouldAddEndPuncttrue
\mciteSetBstMidEndSepPunct{\mcitedefaultmidpunct}
{\mcitedefaultendpunct}{\mcitedefaultseppunct}\relax
\EndOfBibitem
\bibitem[Bi \latin{et~al.}(2017)Bi, Zheng, Chen, Wei, and Huang]{Bi_1400_2017}
Bi,~C.; Zheng,~X.; Chen,~B.; Wei,~H.; Huang,~J. {Spontaneous Passivation of
  Hybrid Perovskite by Sodium Ions from Glass Substrates: Mysterious
  Enhancement of Device Efficiency Revealed}. \emph{ACS Energy Lett.}
  \textbf{2017}, \emph{2}, 1400--1406, DOI:
  \doi{10.1021/acsenergylett.7b00356}\relax
\mciteBstWouldAddEndPuncttrue
\mciteSetBstMidEndSepPunct{\mcitedefaultmidpunct}
{\mcitedefaultendpunct}{\mcitedefaultseppunct}\relax
\EndOfBibitem
\bibitem[Li \latin{et~al.}(2020)Li, Wu, Qi, Zhou, Li, Cheng, Zhao, Li, and
  Zhang]{Li_105237_2020}
Li,~Y.; Wu,~H.; Qi,~W.; Zhou,~X.; Li,~J.; Cheng,~J.; Zhao,~Y.; Li,~Y.;
  Zhang,~X. Passivation of Defects in Perovskite Solar Cell: From a Chemistry
  Point of View. \emph{Nano Energy} \textbf{2020}, \emph{77}, 105237, DOI:
  \doi{10.1016/j.nanoen.2020.105237}\relax
\mciteBstWouldAddEndPuncttrue
\mciteSetBstMidEndSepPunct{\mcitedefaultmidpunct}
{\mcitedefaultendpunct}{\mcitedefaultseppunct}\relax
\EndOfBibitem
\bibitem[Zhang \latin{et~al.}(2018)Zhang, Kim, and Zhu]{Zhang_105_2018}
Zhang,~F.; Kim,~D.~H.; Zhu,~K. 3D/2D Multidimensional Perovskites: Balance of
  High Performance and Stability for Perovskite Solar Cells. \emph{Current
  Opinion in Electrochemistry} \textbf{2018}, \emph{11}, 105--113, DOI:
  \doi{10.1016/j.coelec.2018.10.001}\relax
\mciteBstWouldAddEndPuncttrue
\mciteSetBstMidEndSepPunct{\mcitedefaultmidpunct}
{\mcitedefaultendpunct}{\mcitedefaultseppunct}\relax
\EndOfBibitem
\bibitem[Gangadharan and Ma(2019)Gangadharan, and
  Ma]{ThrithamarasseryGangadharan_2860_2019}
Gangadharan,~D.~T.; Ma,~D. Searching for Stability at Lower Dimensions: Current
  Trends and Future Prospects of Layered Perovskite Solar Cells. \emph{Energy
  {\&} Environmental Science} \textbf{2019}, \emph{12}, 2860--2889, DOI:
  \doi{10.1039/c9ee01591d}\relax
\mciteBstWouldAddEndPuncttrue
\mciteSetBstMidEndSepPunct{\mcitedefaultmidpunct}
{\mcitedefaultendpunct}{\mcitedefaultseppunct}\relax
\EndOfBibitem
\bibitem[Lee \latin{et~al.}(2021)Lee, Kumar, Tyagi, He, Sahani, and
  Kang]{Lee_100759_2021}
Lee,~H.~B.; Kumar,~N.; Tyagi,~B.; He,~S.; Sahani,~R.; Kang,~J.-W. {Bulky
  Organic Cations Engineered Lead-Halide Perovskites: a Review on
  Dimensionality and Optoelectronic Applications}. \emph{Mater. Today Energy}
  \textbf{2021}, \emph{21}, 100759, DOI:
  \doi{10.1016/j.mtener.2021.100759}\relax
\mciteBstWouldAddEndPuncttrue
\mciteSetBstMidEndSepPunct{\mcitedefaultmidpunct}
{\mcitedefaultendpunct}{\mcitedefaultseppunct}\relax
\EndOfBibitem
\bibitem[Ruddlesden and Popper(1957)Ruddlesden, and
  Popper]{Ruddlesden_538_1957}
Ruddlesden,~S.~N.; Popper,~P. New Compounds of the \ce{K2NIF4} type. \emph{Acta
  Crystallographica} \textbf{1957}, \emph{10}, 538--539, DOI:
  \doi{10.1107/s0365110x57001929}\relax
\mciteBstWouldAddEndPuncttrue
\mciteSetBstMidEndSepPunct{\mcitedefaultmidpunct}
{\mcitedefaultendpunct}{\mcitedefaultseppunct}\relax
\EndOfBibitem
\bibitem[G{\'{e}}lvez-Rueda \latin{et~al.}(2020)G{\'{e}}lvez-Rueda, Gompel,
  Herckens, Lutsen, Vanderzande, and Grozema]{GlvezRueda_824_2020}
G{\'{e}}lvez-Rueda,~M.~C.; Gompel,~W. T. M.~V.; Herckens,~R.; Lutsen,~L.;
  Vanderzande,~D.; Grozema,~F.~C. Inducing Charge Separation in Solid-State
  Two-Dimensional Hybrid Perovskites through the Incorporation of Organic
  Charge-Transfer Complexes. \emph{The Journal of Physical Chemistry Letters}
  \textbf{2020}, \emph{11}, 824--830, DOI:
  \doi{10.1021/acs.jpclett.9b03746}\relax
\mciteBstWouldAddEndPuncttrue
\mciteSetBstMidEndSepPunct{\mcitedefaultmidpunct}
{\mcitedefaultendpunct}{\mcitedefaultseppunct}\relax
\EndOfBibitem
\bibitem[Zhou \latin{et~al.}(2019)Zhou, Huang, Sun, Zhang, Li, Lun, Wang, Hong,
  Chen, and Zhou]{Zhou_1901566_2019}
Zhou,~N.; Huang,~B.; Sun,~M.; Zhang,~Y.; Li,~L.; Lun,~Y.; Wang,~X.; Hong,~J.;
  Chen,~Q.; Zhou,~H. The Spacer Cations Interplay for Efficient and Stable
  Layered 2D Perovskite Solar Cells. \emph{Advanced Energy Materials}
  \textbf{2019}, \emph{10}, 1901566, DOI: \doi{10.1002/aenm.201901566}\relax
\mciteBstWouldAddEndPuncttrue
\mciteSetBstMidEndSepPunct{\mcitedefaultmidpunct}
{\mcitedefaultendpunct}{\mcitedefaultseppunct}\relax
\EndOfBibitem
\bibitem[Li \latin{et~al.}(2023)Li, Yan, Cao, Liang, Zhu, Peng, Yang, Liang,
  Zhao, Zang, Zhang, and Song]{Li_e202217910_2023}
Li,~P.; Yan,~L.; Cao,~Q.; Liang,~C.; Zhu,~H.; Peng,~S.; Yang,~Y.; Liang,~Y.;
  Zhao,~R.; Zang,~S.; Zhang,~Y.; Song,~Y. {Dredging the Charge-Carrier Transfer
  Pathway for Efficient Low-Dimensional Ruddlesden-Popper Perovskite Solar
  Cells}. \emph{Angew. Chem. Int. Ed.} \textbf{2023}, \emph{62}, e202217910,
  DOI: \doi{10.1002/anie.202217910}\relax
\mciteBstWouldAddEndPuncttrue
\mciteSetBstMidEndSepPunct{\mcitedefaultmidpunct}
{\mcitedefaultendpunct}{\mcitedefaultseppunct}\relax
\EndOfBibitem
\bibitem[Tsai \latin{et~al.}(2020)Tsai, Liu, Shrestha, Fernando, Tretiak,
  Scott, Vo, Strzalka, and Nie]{Tsai_eaay0815_2020}
Tsai,~H.; Liu,~F.; Shrestha,~S.; Fernando,~K.; Tretiak,~S.; Scott,~B.;
  Vo,~D.~T.; Strzalka,~J.; Nie,~W. A Sensitive and Robust Thin-Film X-ray
  Detector Using 2D Layered Perovskite Diodes. \emph{Science Advances}
  \textbf{2020}, \emph{6}, eaay0815, DOI: \doi{10.1126/sciadv.aay0815}\relax
\mciteBstWouldAddEndPuncttrue
\mciteSetBstMidEndSepPunct{\mcitedefaultmidpunct}
{\mcitedefaultendpunct}{\mcitedefaultseppunct}\relax
\EndOfBibitem
\bibitem[Wang \latin{et~al.}(2020)Wang, Wang, Chen, Zhang, Jiang, Zhang, Qin,
  Wang, Li, Pan, Liu, Shi, Zhang, Tu, Wang, Long, Li, Lin, Wang, Zhan, Shen,
  Meng, and Chu]{Wang_1901402_2020}
Wang,~H.; Wang,~X.; Chen,~Y.; Zhang,~S.; Jiang,~W.; Zhang,~X.; Qin,~J.;
  Wang,~J.; Li,~X.; Pan,~Y.; Liu,~F.; Shi,~Z.; Zhang,~H.; Tu,~L.; Wang,~H.;
  Long,~H.; Li,~D.; Lin,~T.; Wang,~J.; Zhan,~Y.; Shen,~H.; Meng,~X.; Chu,~J.
  Extremely Low Dark Current \ce{MoS2} Photodetector via 2D Halide Perovskite
  as the Electron Reservoir. \emph{Advanced Optical Materials} \textbf{2020},
  \emph{8}, 1901402, DOI: \doi{10.1002/adom.201901402}\relax
\mciteBstWouldAddEndPuncttrue
\mciteSetBstMidEndSepPunct{\mcitedefaultmidpunct}
{\mcitedefaultendpunct}{\mcitedefaultseppunct}\relax
\EndOfBibitem
\bibitem[Stoumpos \latin{et~al.}(2016)Stoumpos, Cao, Clark, Young, Rondinelli,
  Jang, Hupp, and Kanatzidis]{Stoumpos_2852_2016}
Stoumpos,~C.~C.; Cao,~D.~H.; Clark,~D.~J.; Young,~J.; Rondinelli,~J.~M.;
  Jang,~J.~I.; Hupp,~J.~T.; Kanatzidis,~M.~G. {Ruddlesden{\textendash}Popper
  Hybrid Lead Iodide Perovskite 2D Homologous Semiconductors}. \emph{Chem.
  Mater.} \textbf{2016}, \emph{28}, 2852--2867, DOI:
  \doi{10.1021/acs.chemmater.6b00847}\relax
\mciteBstWouldAddEndPuncttrue
\mciteSetBstMidEndSepPunct{\mcitedefaultmidpunct}
{\mcitedefaultendpunct}{\mcitedefaultseppunct}\relax
\EndOfBibitem
\bibitem[Wang \latin{et~al.}(2020)Wang, Xu, Wang, Liu, Chen, Zheng, Ji, Zhang,
  Zhang, Chen, Wu, Chen, and Li]{Wang_123589_2020}
Wang,~Y.; Xu,~H.; Wang,~F.; Liu,~D.; Chen,~H.; Zheng,~H.; Ji,~L.; Zhang,~P.;
  Zhang,~T.; Chen,~Z.~D.; Wu,~J.; Chen,~L.; Li,~S. {Unveiling the Guest Effect
  of N-butylammonium Iodide Towards Efficient and Stable 2D-3D Perovskite Solar
  Cells Through Sequential Deposition Process}. \emph{Chem. Eng. J.}
  \textbf{2020}, \emph{391}, 123589, DOI: \doi{10.1016/j.cej.2019.123589}\relax
\mciteBstWouldAddEndPuncttrue
\mciteSetBstMidEndSepPunct{\mcitedefaultmidpunct}
{\mcitedefaultendpunct}{\mcitedefaultseppunct}\relax
\EndOfBibitem
\bibitem[Qin \latin{et~al.}(2020)Qin, Zhong, Intemann, Leng, Cui, Qin, Xiong,
  Liu, Jen, and Yao]{Qin_1904050_2020}
Qin,~Y.; Zhong,~H.; Intemann,~J.~J.; Leng,~S.; Cui,~M.; Qin,~C.; Xiong,~M.;
  Liu,~F.; Jen,~A. K.-Y.; Yao,~K. {Coordination Engineering of Single-Crystal
  Precursor for Phase Control in Ruddlesden{\textendash}Popper Perovskite Solar
  Cells}. \emph{Adv. Energy Mater.} \textbf{2020}, \emph{10}, 1904050, DOI:
  \doi{10.1002/aenm.201904050}\relax
\mciteBstWouldAddEndPuncttrue
\mciteSetBstMidEndSepPunct{\mcitedefaultmidpunct}
{\mcitedefaultendpunct}{\mcitedefaultseppunct}\relax
\EndOfBibitem
\bibitem[Cohen \latin{et~al.}(2019)Cohen, Li, Meng, and Etgar]{Cohen_2588_2019}
Cohen,~B.-E.; Li,~Y.; Meng,~Q.; Etgar,~L. {Dion{\textendash}Jacobson
  Two-Dimensional Perovskite Solar Cells Based on Benzene Dimethanammonium
  Cation}. \emph{Nano Lett.} \textbf{2019}, \emph{19}, 2588--2597, DOI:
  \doi{10.1021/acs.nanolett.9b00387}\relax
\mciteBstWouldAddEndPuncttrue
\mciteSetBstMidEndSepPunct{\mcitedefaultmidpunct}
{\mcitedefaultendpunct}{\mcitedefaultseppunct}\relax
\EndOfBibitem
\bibitem[Lao \latin{et~al.}(2022)Lao, Yang, Yu, Guo, Zou, Chen, and
  Xiao]{Lao_2105307_2022}
Lao,~Y.; Yang,~S.; Yu,~W.; Guo,~H.; Zou,~Y.; Chen,~Z.; Xiao,~L.
  {Multifunctional {$\pi$}-Conjugated Additives for Halide Perovskite}.
  \emph{Adv. Sci.} \textbf{2022}, \emph{9}, 2105307, DOI:
  \doi{10.1002/advs.202105307}\relax
\mciteBstWouldAddEndPuncttrue
\mciteSetBstMidEndSepPunct{\mcitedefaultmidpunct}
{\mcitedefaultendpunct}{\mcitedefaultseppunct}\relax
\EndOfBibitem
\bibitem[Hohenberg and Kohn(1964)Hohenberg, and Kohn]{Hohenberg_B864_1964}
Hohenberg,~P.; Kohn,~W. Inhomogeneous Electron Gas. \emph{Phys. Rev.}
  \textbf{1964}, \emph{136}, B864--B871, DOI:
  \doi{10.1103/PhysRev.136.B864}\relax
\mciteBstWouldAddEndPuncttrue
\mciteSetBstMidEndSepPunct{\mcitedefaultmidpunct}
{\mcitedefaultendpunct}{\mcitedefaultseppunct}\relax
\EndOfBibitem
\bibitem[Kohn and Sham(1965)Kohn, and Sham]{Kohn_A1133_1965}
Kohn,~W.; Sham,~L.~J. Self-consistent Equations Including Exchange and
  Correlation Effects. \emph{Phys. Rev.} \textbf{1965}, \emph{140},
  A1133--A1138, DOI: \doi{10.1103/PhysRev.140.A1133}\relax
\mciteBstWouldAddEndPuncttrue
\mciteSetBstMidEndSepPunct{\mcitedefaultmidpunct}
{\mcitedefaultendpunct}{\mcitedefaultseppunct}\relax
\EndOfBibitem
\bibitem[Perdew \latin{et~al.}(1996)Perdew, Burke, and
  Ernzerhof]{Perdew_3865_1996}
Perdew,~J.~P.; Burke,~K.; Ernzerhof,~M. Generalized Gradient Approximation Made
  Simple. \emph{Phys. Rev. Lett.} \textbf{1996}, \emph{77}, 3865--3868, DOI:
  \doi{10.1103/PhysRevLett.77.3865}\relax
\mciteBstWouldAddEndPuncttrue
\mciteSetBstMidEndSepPunct{\mcitedefaultmidpunct}
{\mcitedefaultendpunct}{\mcitedefaultseppunct}\relax
\EndOfBibitem
\bibitem[Kresse and Hafner(1993)Kresse, and Hafner]{Kresse_558_1993}
Kresse,~G.; Hafner,~J. Ab Initio Molecular Dynamics for Liquid Metals.
  \emph{Phys. Rev. B} \textbf{1993}, \emph{47}, 558--561(R), DOI:
  \doi{10.1103/PhysRevB.47.558}\relax
\mciteBstWouldAddEndPuncttrue
\mciteSetBstMidEndSepPunct{\mcitedefaultmidpunct}
{\mcitedefaultendpunct}{\mcitedefaultseppunct}\relax
\EndOfBibitem
\bibitem[Kresse and Furthm{\"u}ller(1996)Kresse, and
  Furthm{\"u}ller]{Kresse_11169_1996}
Kresse,~G.; Furthm{\"u}ller,~J. Efficient Iterative Schemes for
  \textit{Ab{~}initio} Total-energy Calculations Using a Plane-wave Basis set.
  \emph{Phys. Rev. B} \textbf{1996}, \emph{54}, 11169--11186, DOI:
  \doi{10.1103/physrevb.54.11169}\relax
\mciteBstWouldAddEndPuncttrue
\mciteSetBstMidEndSepPunct{\mcitedefaultmidpunct}
{\mcitedefaultendpunct}{\mcitedefaultseppunct}\relax
\EndOfBibitem
\bibitem[Blochl(1994)]{Blochl_17953_1994}
Blochl,~P.~E. Projector Augmented-wave Method. \emph{Phys. Rev. B}
  \textbf{1994}, \emph{50}, 17953--17979, DOI:
  \doi{10.1103/PhysRevB.50.17953}\relax
\mciteBstWouldAddEndPuncttrue
\mciteSetBstMidEndSepPunct{\mcitedefaultmidpunct}
{\mcitedefaultendpunct}{\mcitedefaultseppunct}\relax
\EndOfBibitem
\bibitem[Kresse and Joubert(1999)Kresse, and Joubert]{Kresse_1758_1999}
Kresse,~G.; Joubert,~D. From Ultrasoft Pseudopotentials to the Projector
  Augmented-wave Method. \emph{Phys. Rev. B} \textbf{1999}, \emph{59},
  1758--1775, DOI: \doi{10.1103/PhysRevB.59.1758}\relax
\mciteBstWouldAddEndPuncttrue
\mciteSetBstMidEndSepPunct{\mcitedefaultmidpunct}
{\mcitedefaultendpunct}{\mcitedefaultseppunct}\relax
\EndOfBibitem
\bibitem[Moraes \latin{et~al.}(2023)Moraes, Bittencourt, Andriani, and
  Da~Silva]{Moraes_16357_2023}
Moraes,~P. I.~R.; Bittencourt,~A. F.~B.; Andriani,~K.~F.; Da~Silva,~J. L.~F.
  {Theoretical Insights into Methane Activation on Transition-Metal Single-Atom
  Catalysts Supported on the CeO2(111) Surface}. \emph{J. Phys. Chem. C}
  \textbf{2023}, \emph{127}, 16357--16366, DOI:
  \doi{10.1021/acs.jpcc.3c02653}\relax
\mciteBstWouldAddEndPuncttrue
\mciteSetBstMidEndSepPunct{\mcitedefaultmidpunct}
{\mcitedefaultendpunct}{\mcitedefaultseppunct}\relax
\EndOfBibitem
\bibitem[Grimme \latin{et~al.}(2010)Grimme, Antony, Ehrlich, and
  Krieg]{Grimme_154104_2010}
Grimme,~S.; Antony,~J.; Ehrlich,~S.; Krieg,~H. A Consistent and Accurate Ab
  Initio Parametrization of Density Functional Dispersion Correction (dft-d)
  for the 94 Elements H-pu. \emph{J. Chem. Phys.} \textbf{2010}, \emph{132},
  154104, DOI: \doi{10.1063/1.3382344}\relax
\mciteBstWouldAddEndPuncttrue
\mciteSetBstMidEndSepPunct{\mcitedefaultmidpunct}
{\mcitedefaultendpunct}{\mcitedefaultseppunct}\relax
\EndOfBibitem
\bibitem[Oz{\'o}rio \latin{et~al.}(2020)Oz{\'o}rio, Oliveira, Silveira,
  Nogueira, and Da{~}Silva]{Ozorio_3439_2020}
Oz{\'o}rio,~M.~S.; Oliveira,~W. X.~C.; Silveira,~J. F. R.~V.; Nogueira,~A.~F.;
  Da{~}Silva,~J. L.~F. Novel Zero-dimensional Lead-free Bismuth Based
  Perovskites: From Synthesis to Structural and Optoelectronic
  Characterization. \emph{Mater. Adv.} \textbf{2020}, \emph{1}, 3439--3448,
  DOI: \doi{10.1039/d0ma00791a}\relax
\mciteBstWouldAddEndPuncttrue
\mciteSetBstMidEndSepPunct{\mcitedefaultmidpunct}
{\mcitedefaultendpunct}{\mcitedefaultseppunct}\relax
\EndOfBibitem
\bibitem[Francis and Payne(1990)Francis, and Payne]{Francis_4395_1990}
Francis,~G.~P.; Payne,~M.~C. Finite Basis Set Corrections to Total Energy
  Pseudopotential Calculations. \emph{Journal of Physics: Condensed Matter}
  \textbf{1990}, \emph{2}, 4395--4404, DOI:
  \doi{10.1088/0953-8984/2/19/007}\relax
\mciteBstWouldAddEndPuncttrue
\mciteSetBstMidEndSepPunct{\mcitedefaultmidpunct}
{\mcitedefaultendpunct}{\mcitedefaultseppunct}\relax
\EndOfBibitem
\bibitem[Manz and Limas(2016)Manz, and Limas]{Manz_47771_2016}
Manz,~T.~A.; Limas,~N.~G. {Introducing DDEC6 Atomic Population Analysis: Part
  1. Charge Partitioning Theory and Methodology}. \emph{RSC Adv.}
  \textbf{2016}, \emph{6}, 47771--47801, DOI: \doi{10.1039/C6RA04656H}\relax
\mciteBstWouldAddEndPuncttrue
\mciteSetBstMidEndSepPunct{\mcitedefaultmidpunct}
{\mcitedefaultendpunct}{\mcitedefaultseppunct}\relax
\EndOfBibitem
\bibitem[Limas and Manz(2016)Limas, and Manz]{Limas_45727_2016}
Limas,~N.~G.; Manz,~T.~A. {Introducing DDEC6 atomic Population Analysis: Part
  2. Computed Results for a Wide Range of Periodic and Nonperiodic Materials}.
  \emph{RSC Adv.} \textbf{2016}, \emph{6}, 45727--45747, DOI:
  \doi{10.1039/C6RA05507A}\relax
\mciteBstWouldAddEndPuncttrue
\mciteSetBstMidEndSepPunct{\mcitedefaultmidpunct}
{\mcitedefaultendpunct}{\mcitedefaultseppunct}\relax
\EndOfBibitem
\bibitem[Marchenko \latin{et~al.}(2020)Marchenko, Fateev, Petrov, Korolev,
  Mitrofanov, Petrov, Goodilin, and Tarasov]{Marchenko_7383_2020}
Marchenko,~E.~I.; Fateev,~S.~A.; Petrov,~A.~A.; Korolev,~V.~V.; Mitrofanov,~A.;
  Petrov,~A.~V.; Goodilin,~E.~A.; Tarasov,~A.~B. Database of Two-Dimensional
  Hybrid Perovskite Materials: Open-Access Collection of Crystal Structures,
  Band Gaps, and Atomic Partial Charges Predicted by Machine Learning.
  \emph{Chemistry of Materials} \textbf{2020}, \emph{32}, 7383--7388, DOI:
  \doi{10.1021/acs.chemmater.0c02290}\relax
\mciteBstWouldAddEndPuncttrue
\mciteSetBstMidEndSepPunct{\mcitedefaultmidpunct}
{\mcitedefaultendpunct}{\mcitedefaultseppunct}\relax
\EndOfBibitem
\bibitem[Moral \latin{et~al.}(2020)Moral, Germino, Bonato, Almeida,
  Ther{\'{e}}zio, Atvars, Stranks, Nome, and Nogueira]{Moral_2001431_2020}
Moral,~R.~F.; Germino,~J.~C.; Bonato,~L.~G.; Almeida,~D.~B.;
  Ther{\'{e}}zio,~E.~M.; Atvars,~T. D.~Z.; Stranks,~S.~D.; Nome,~R.~A.;
  Nogueira,~A.~F. Influence of the Vibrational Modes from the Organic Moieties
  in 2D Lead Halides on Excitonic Recombination and Phase Transition.
  \emph{Advanced Optical Materials} \textbf{2020}, \emph{8}, 2001431, DOI:
  \doi{10.1002/adom.202001431}\relax
\mciteBstWouldAddEndPuncttrue
\mciteSetBstMidEndSepPunct{\mcitedefaultmidpunct}
{\mcitedefaultendpunct}{\mcitedefaultseppunct}\relax
\EndOfBibitem
\bibitem[Zhang \latin{et~al.}(2020)Zhang, Lu, Tong, Berry, Beard, and
  Zhu]{Zhang_1154_2020}
Zhang,~F.; Lu,~H.; Tong,~J.; Berry,~J.~J.; Beard,~M.~C.; Zhu,~K. Advances in
  Two-Dimensional Organic{\textendash}Inorganic Hybrid Perovskites.
  \emph{Energy Environ. Sci.} \textbf{2020}, \emph{13}, 1154--1186, DOI:
  \doi{10.1039/c9ee03757h}\relax
\mciteBstWouldAddEndPuncttrue
\mciteSetBstMidEndSepPunct{\mcitedefaultmidpunct}
{\mcitedefaultendpunct}{\mcitedefaultseppunct}\relax
\EndOfBibitem
\bibitem[Hou \latin{et~al.}(2020)Hou, Wu, Yang, Ye, Honavar, van Duin, Wang,
  and Priya]{Hou_060906_2020}
Hou,~Y.; Wu,~C.; Yang,~D.; Ye,~T.; Honavar,~V.~G.; van Duin,~A. C.~T.;
  Wang,~K.; Priya,~S. Two-Dimensional Hybrid Organic{\textendash}Inorganic
  Perovskites as Emergent Ferroelectric Materials. \emph{Journal of Applied
  Physics} \textbf{2020}, \emph{128}, 060906, DOI:
  \doi{10.1063/5.0016010}\relax
\mciteBstWouldAddEndPuncttrue
\mciteSetBstMidEndSepPunct{\mcitedefaultmidpunct}
{\mcitedefaultendpunct}{\mcitedefaultseppunct}\relax
\EndOfBibitem
\bibitem[Da{~}Silva \latin{et~al.}(2006)Da{~}Silva, Stampfl, and
  Scheffler]{DaSilva_703_2006}
Da{~}Silva,~J. L.~F.; Stampfl,~C.; Scheffler,~M. Converged Properties of Clean
  Metal Surfaces by All-electron First-principles Calculations. \emph{Surf.
  Sci.} \textbf{2006}, \emph{600}, 703--715, DOI:
  \doi{10.1016/j.susc.2005.12.008}\relax
\mciteBstWouldAddEndPuncttrue
\mciteSetBstMidEndSepPunct{\mcitedefaultmidpunct}
{\mcitedefaultendpunct}{\mcitedefaultseppunct}\relax
\EndOfBibitem
\bibitem[Cheng \latin{et~al.}(2018)Cheng, Li, Maity, Wei, Nordlund, Ho, Lien,
  Lin, Liang, Miao, Ajia, Yin, Sokaras, Javey, Roqan, Mohammed, and
  He]{Cheng_1_2018}
Cheng,~B.; Li,~T.-Y.; Maity,~P.; Wei,~P.-C.; Nordlund,~D.; Ho,~K.-T.;
  Lien,~D.-H.; Lin,~C.-H.; Liang,~R.-Z.; Miao,~X.; Ajia,~I.~A.; Yin,~J.;
  Sokaras,~D.; Javey,~A.; Roqan,~I.~S.; Mohammed,~O.~F.; He,~J.-H. {Extremely
  Reduced Dielectric Confinement in Two-Dimensional Hybrid Perovskites with
  Large Polar Organics}. \emph{Commun. Phys.} \textbf{2018}, \emph{1}, 1--8,
  DOI: \doi{10.1038/s42005-018-0082-8}\relax
\mciteBstWouldAddEndPuncttrue
\mciteSetBstMidEndSepPunct{\mcitedefaultmidpunct}
{\mcitedefaultendpunct}{\mcitedefaultseppunct}\relax
\EndOfBibitem
\bibitem[Oz{\ifmmode\acute{o}\else\'{o}\fi}rio
  \latin{et~al.}(2021)Oz{\ifmmode\acute{o}\else\'{o}\fi}rio, Srikanth, Besse,
  and Da~Silva]{Ozorio_2286_2021}
Oz{\ifmmode\acute{o}\else\'{o}\fi}rio,~M.~S.; Srikanth,~M.; Besse,~R.;
  Da~Silva,~J. L.~F. {The Role of the A-cations in the Polymorphic Stability
  and Optoelectronic Properties of Lead-Free \ce{ASnI3} Perovskites}.
  \emph{Phys. Chem. Chem. Phys.} \textbf{2021}, \emph{23}, 2286--2297, DOI:
  \doi{10.1039/D0CP06090A}\relax
\mciteBstWouldAddEndPuncttrue
\mciteSetBstMidEndSepPunct{\mcitedefaultmidpunct}
{\mcitedefaultendpunct}{\mcitedefaultseppunct}\relax
\EndOfBibitem
\bibitem[Leung \latin{et~al.}(2022)Leung, Ahmad, Syed, and Ng]{Leung_1_2022}
Leung,~T.~L.; Ahmad,~I.; Syed,~A.~A.; Ng,~A.~B.,~Alan Man~Ching {Stability of
  2D and quasi-2D Perovskite Materials and Devices}. \emph{Commun. Mater.}
  \textbf{2022}, \emph{3}, 1--10, DOI: \doi{10.1038/s43246-022-00285-9}\relax
\mciteBstWouldAddEndPuncttrue
\mciteSetBstMidEndSepPunct{\mcitedefaultmidpunct}
{\mcitedefaultendpunct}{\mcitedefaultseppunct}\relax
\EndOfBibitem
\bibitem[Espinosa \latin{et~al.}(1998)Espinosa, Molins, and
  Lecomte]{Espinosa_170_1998}
Espinosa,~E.; Molins,~E.; Lecomte,~C. {Hydrogen Bond Strengths Revealed by
  Topological Analyses of Experimentally Observed Electron Densities}.
  \textbf{1998}, \emph{285}, 170--173, DOI:
  \doi{10.1016/S0009-2614(98)00036-0}\relax
\mciteBstWouldAddEndPuncttrue
\mciteSetBstMidEndSepPunct{\mcitedefaultmidpunct}
{\mcitedefaultendpunct}{\mcitedefaultseppunct}\relax
\EndOfBibitem
\bibitem[Cuthriell \latin{et~al.}(2023)Cuthriell, Malliakas, Kanatzidis, and
  Schaller]{Cuthriell_11710_2023}
Cuthriell,~S.~A.; Malliakas,~C.~D.; Kanatzidis,~M.~G.; Schaller,~R.~D. {Cyclic
  versus Linear Alkylammonium Cations: Preventing Phase Transitions at
  Operational Temperatures in 2D Perovskites}. \emph{J. Am. Chem. Soc.}
  \textbf{2023}, \emph{145}, 11710--11716, DOI:
  \doi{10.1021/jacs.3c02008}\relax
\mciteBstWouldAddEndPuncttrue
\mciteSetBstMidEndSepPunct{\mcitedefaultmidpunct}
{\mcitedefaultendpunct}{\mcitedefaultseppunct}\relax
\EndOfBibitem
\bibitem[Gao \latin{et~al.}(2023)Gao, Dong, and Liu]{Gao_1_2023}
Gao,~Y.; Dong,~X.; Liu,~Y. {Recent Progress of Layered Perovskite Solar Cells
  Incorporating Aromatic Spacers}. \emph{Nano-Micro Lett.} \textbf{2023},
  \emph{15}, 1--19, DOI: \doi{10.1007/s40820-023-01141-2}\relax
\mciteBstWouldAddEndPuncttrue
\mciteSetBstMidEndSepPunct{\mcitedefaultmidpunct}
{\mcitedefaultendpunct}{\mcitedefaultseppunct}\relax
\EndOfBibitem
\bibitem[Temerova \latin{et~al.}(2022)Temerova, Chou, Kisel, Eskelinen,
  Kinnunen, J{\ifmmode\ddot{a}\else\"{a}\fi}nis, Karttunen, Chou, and
  Koshevoy]{Temerova_19220_2022}
Temerova,~D.; Chou,~T.-C.; Kisel,~K.~S.; Eskelinen,~T.; Kinnunen,~N.;
  J{\ifmmode\ddot{a}\else\"{a}\fi}nis,~J.; Karttunen,~A.~J.; Chou,~P.-T.;
  Koshevoy,~I.~O. {Hybrid Inorganic{\textendash}Organic Complexes of Zn, Cd,
  and Pb with a Cationic Phenanthro-diimine Ligand}. \emph{Inorg. Chem.}
  \textbf{2022}, \emph{61}, 19220--19231, DOI:
  \doi{10.1021/acs.inorgchem.2c02867}\relax
\mciteBstWouldAddEndPuncttrue
\mciteSetBstMidEndSepPunct{\mcitedefaultmidpunct}
{\mcitedefaultendpunct}{\mcitedefaultseppunct}\relax
\EndOfBibitem
\bibitem[Jahn~H. and Teller(1937)Jahn~H., and Teller]{JahnH._220_1937}
Jahn~H.,~A.; Teller,~E. Stability of Polyatomic Molecules in Degenerate
  Electronic States - I-orbital Degeneracy. \emph{Proc. R. Soc. London A -
  Math. Phys. Sci.} \textbf{1937}, \emph{161}, 220--235, DOI:
  \doi{10.1098/rspa.1937.0142}\relax
\mciteBstWouldAddEndPuncttrue
\mciteSetBstMidEndSepPunct{\mcitedefaultmidpunct}
{\mcitedefaultendpunct}{\mcitedefaultseppunct}\relax
\EndOfBibitem
\bibitem[Regis \latin{et~al.}(2024)Regis, Da~Silva, and
  Lima]{Regis_107710_2024}
Regis,~N.~M.; Da~Silva,~J. L.~F.; Lima,~M.~P. {Ab initio Investigation of the
  Adsorption Properties of Molecules on \ce{MoS2} Pristine and ith Sulfur
  Vacancy}. \emph{Mater. Today Commun.} \textbf{2024}, \emph{38}, 107710, DOI:
  \doi{10.1016/j.mtcomm.2023.107710}\relax
\mciteBstWouldAddEndPuncttrue
\mciteSetBstMidEndSepPunct{\mcitedefaultmidpunct}
{\mcitedefaultendpunct}{\mcitedefaultseppunct}\relax
\EndOfBibitem
\bibitem[Mitzi \latin{et~al.}(1995)Mitzi, Wang, Feild, Chess, and
  Guloy]{Mitzi_1473_1995}
Mitzi,~D.~B.; Wang,~S.; Feild,~C.~A.; Chess,~C.~A.; Guloy,~A.~M. {Conducting
  Layered Organic-inorganic Halides Containing <110>-Oriented Perovskite
  Sheets}. \emph{Science} \textbf{1995}, \emph{267}, 1473--1476, DOI:
  \doi{10.1126/science.267.5203.1473}\relax
\mciteBstWouldAddEndPuncttrue
\mciteSetBstMidEndSepPunct{\mcitedefaultmidpunct}
{\mcitedefaultendpunct}{\mcitedefaultseppunct}\relax
\EndOfBibitem
\bibitem[Leng \latin{et~al.}(2020)Leng, Wang, Shao, Abdelwahab, Grinblat,
  Verzhbitskiy, Li, Cai, Chi, Fu, Song, Rusydi, Eda, Maier, and
  Loh]{Leng_2041_2020}
Leng,~K.; Wang,~L.; Shao,~Y.; Abdelwahab,~I.; Grinblat,~G.; Verzhbitskiy,~I.;
  Li,~R.; Cai,~Y.; Chi,~X.; Fu,~W.; Song,~P.; Rusydi,~A.; Eda,~G.;
  Maier,~S.~A.; Loh,~K.~P. {Electron Tunneling at the Molecularly Thin 2D
  Perovskite and Graphene van der Waals Interface}. \emph{Nat. Commun.}
  \textbf{2020}, \emph{11}, 1--8, DOI: \doi{10.1038/s41467-020-19331-6}\relax
\mciteBstWouldAddEndPuncttrue
\mciteSetBstMidEndSepPunct{\mcitedefaultmidpunct}
{\mcitedefaultendpunct}{\mcitedefaultseppunct}\relax
\EndOfBibitem
\bibitem[Zheng \latin{et~al.}(2021)Zheng, Fang, Shang, Sun, Zheng, Yang, Hou,
  and Yang]{Zheng_2328_2021}
Zheng,~Y.; Fang,~Z.; Shang,~M.; Sun,~Q.; Zheng,~J.; Yang,~Z.; Hou,~X.; Yang,~W.
  {Linearly Tailored Work Function of Orthorhombic \ce{CsSnI3} Perovskites}.
  \emph{ACS Energy Lett.} \textbf{2021}, \emph{6}, 2328--2335, DOI:
  \doi{10.1021/acsenergylett.1c00977}\relax
\mciteBstWouldAddEndPuncttrue
\mciteSetBstMidEndSepPunct{\mcitedefaultmidpunct}
{\mcitedefaultendpunct}{\mcitedefaultseppunct}\relax
\EndOfBibitem
\bibitem[Liu \latin{et~al.}(2016)Liu, Xiao, and Goddard]{Liu_3335_2016}
Liu,~Y.; Xiao,~H.; Goddard,~W. A.~I. {Two-Dimensional Halide Perovskites:
  Tuning Electronic Activities of Defects}. \emph{Nano Lett.} \textbf{2016},
  \emph{16}, 3335--3340, DOI: \doi{10.1021/acs.nanolett.6b00964}\relax
\mciteBstWouldAddEndPuncttrue
\mciteSetBstMidEndSepPunct{\mcitedefaultmidpunct}
{\mcitedefaultendpunct}{\mcitedefaultseppunct}\relax
\EndOfBibitem
\bibitem[Ye \latin{et~al.}(2022)Ye, Cai, Sun, Xu, Ni, Li, and
  Zhang]{Ye_1566_2022}
Ye,~X.; Cai,~H.; Sun,~Q.; Xu,~T.; Ni,~J.; Li,~J.; Zhang,~J. {Alkylammonium Salt
  with Different Chain Length for High-Efficiency and Good-Stability 2D/3D
  Hybrid Perovskite Solar Cells}. \emph{Org. Electron.} \textbf{2022},
  \emph{106}, 106542, DOI: \doi{10.1016/j.orgel.2022.106542}\relax
\mciteBstWouldAddEndPuncttrue
\mciteSetBstMidEndSepPunct{\mcitedefaultmidpunct}
{\mcitedefaultendpunct}{\mcitedefaultseppunct}\relax
\EndOfBibitem
\bibitem[Li \latin{et~al.}(2021)Li, Dong, Lv, Liu, Lu, Zheng, Dong, Xu, Xie,
  and Liu]{Li_2072_2021}
Li,~Q.; Dong,~Y.; Lv,~G.; Liu,~T.; Lu,~D.; Zheng,~N.; Dong,~X.; Xu,~Z.;
  Xie,~Z.; Liu,~Y. {Fluorinated Aromatic Formamidinium Spacers Boost Efficiency
  of Layered Ruddlesden{\textendash}Popper Perovskite Solar Cells}. \emph{ACS
  Energy Lett.} \textbf{2021}, \emph{6}, 2072--2080, DOI:
  \doi{10.1021/acsenergylett.1c00620}\relax
\mciteBstWouldAddEndPuncttrue
\mciteSetBstMidEndSepPunct{\mcitedefaultmidpunct}
{\mcitedefaultendpunct}{\mcitedefaultseppunct}\relax
\EndOfBibitem
\end{mcitethebibliography}

\end{document}